\documentclass[10pt,usletter,twocolumn]{article}
\usepackage[top=1.85cm, bottom=1.85cm, left=1.85cm, right=1.85cm]{geometry}

\usepackage{amsmath,amssymb,amsfonts}
\usepackage{listings}                       %
\usepackage[noend]{algpseudocode}
\usepackage{multirow}
\usepackage{graphicx}		%
	\setkeys{Gin}{width=\textwidth, height=\textheight, keepaspectratio}
\usepackage{pifont}
\usepackage{textcomp}
\usepackage{xcolor}
\usepackage{flushend}
\usepackage{algorithm}

\usepackage[small,bf]{caption}
\usepackage[square,sort,comma,numbers]{natbib}

\usepackage{xcolor}
\usepackage{booktabs}
\usepackage{paralist}
\usepackage{threeparttable}

\usepackage{times}

\usepackage{titlesec}
\titlespacing*{\section}{0pt}{*3}{3pt}
\titlespacing{\subsection}{0pt}{*2}{2pt}
\titlespacing{\subsubsection}{0pt}{*1}{1pt}

\newcommand{\descr}[1]{\smallskip\noindent\textbf{#1}}

\definecolor{providercol}{HTML}{2b8a3e}
\definecolor{adversarycol}{HTML}{c92a2a}
\definecolor{linkcol}{rgb}{0,0,0.5}
\definecolor{citecol}{rgb}{0,0.5,0.3}
\definecolor{urlcol}{rgb}{0.3,0,0}

\usepackage{xspace}

\renewenvironment{thebibliography}[1]{
  \begin{oldthebibliography}{#1}
    \setlength{\itemsep}{0.0em}
    \setlength{\parskip}{0.0em}
}
{
  \end{oldthebibliography}
}

\usepackage[hang,flushmargin]{footmisc}

\renewcommand{\footnoterule}{%
  \kern -3pt
  \hrule width 1in
  \kern 2pt
}

\usepackage[hyphens]{url}

\makeatletter
\def\url@leostyle{%
  \@ifundefined{selectfont}{\def\UrlFont{}}%
  {\def\UrlFont{}}%
}
\makeatother
\usepackage[hyphenbreaks]{breakurl}

\definecolor{darkred}{RGB}{153,0,0}
\definecolor{darkblue}{RGB}{0,0,99}
\usepackage[colorlinks=true, linkcolor=darkred, citecolor=darkred, urlcolor=darkblue]{hyperref}

\usepackage[aboveskip=2pt,belowskip=-2pt]{subcaption}

\newcommand{\revision}[1]{\textcolor{black}{#1}}
\newcommand{\minor}[1]{\textcolor{black}{#1}}
\newcommand{\myspace}{\vspace{3pt}}
\usepackage{numprint}
\npthousandsep{,}

\title{\bf {\em From HODL to MOON:} Understanding Community Evolution, Emotional Dynamics, and Price Interplay in the Cryptocurrency Ecosystem}

\author{Kostantinos Papadamou$^1$,
Jay Patel$^2$,
Jeremy Blackburn$^2$,
Philipp Jovanovic$^1$,
Emiliano De Cristofaro$^3$\\[0.5ex]
$^1$University College London, 
$^2$Binghamton University,
$^3$UC Riverside
}

\date{}

\begin{document}

\maketitle

\sloppy

\begin{abstract}
This paper presents a large-scale analysis of the cryptocurrency community on Reddit, shedding light on the intricate relationship between the evolution of their activity, emotional dynamics, and price movements. 
\minor{We analyze over 130M posts on 122 cryptocurrency-related subreddits using temporal analysis, statistical modeling, and emotion detection.} %
While /r/CryptoCurrency and /r/dogecoin are the most active subreddits, we find an overall surge in cryptocurrency-related activity in 2021, followed by a sharp decline.

We also uncover a strong relationship in terms of cross-correlation between online activity and the price of various coins, with the changes in the number of posts mostly leading the price changes.
\revision{Backtesting analysis shows that a straightforward strategy based on the cross-correlation where one buys/sells a coin if the daily number of posts about it is greater/less than the previous would have led to a 3x return on investment.}
Finally, we shed light on the emotional dynamics of the cryptocurrency communities, finding that joy becomes a prominent indicator during upward market performance, while a decline in the market manifests an increase in anger.
\end{abstract}

\section{Introduction}\label{sec:intro}

Cryptocurrencies have gained significant popularity in recent years, %
with over \$1 trillion in global market capitalization~\cite{cryptogrowth2021cnbc}.
As the ecosystem attracts more and more people and investments, it becomes increasingly crucial to understand the underlying dynamics that shape its user base and possibly influence market behavior.

A popular, if at times distorted, narrative is that cryptocurrencies operate on decentralized networks that 
empower communities to actively engage in their development and growth. %
As a result, social media plays an important role in facilitating discussions and information sharing among enthusiasts and investors.
These enable them to share opinions and insights and discuss various aspects and events related to cryptocurrencies, including adoption trends and their potential influence on market dynamics~\cite{cryptoelon2021wapo}.

\minor{Prior research studied the growth and characteristics of online discussions surrounding popular cryptocurrencies like Bitcoin and Ethereum, although only on small scales, or for short timeframes, and on a small set of cryptocurrencies~\cite{glenski2019characterizing,linton2017dynamic}\revision{~\cite{phillips2018mutual}}.
At the same time, most of this work focuses on extracting insights to predict cryptocurrency price movements~\cite{kim2016predicting,wooley2019extracting}\revision{~\cite{phillips2017predicting}.}}

However, the impact of communities centered around cryptocurrencies on the overall ecosystem remains relatively understudied, especially at a large scale.
Overall, their intricate mechanisms as well as the factors influencing price fluctuations remain elusive.
As a result, we set out to study the evolution of online cryptocurrency communities and the interplay between their evolution, emotional dynamics, and price movements, %
aiming to shed light on the factors and underlying mechanisms that drive this dynamic ecosystem.

\descr{Research Questions.} With this motivation in mind, this paper explores the footprints of cryptocurrency communities on Reddit, and aims to answer three research questions:\smallskip
\begin{compactitem}
    \item \textbf{RQ1:} How have cryptocurrency communities on Reddit evolved in activity and popularity over the years?
    \item \textbf{RQ2:} How do cryptocurrency price movements affect activity in relevant communities and vice-versa? 
    \item \textbf{RQ3:} How do events and changes in cryptocurrency price correlate with the emotions within cryptocurrency-related communities' discussions?\smallskip
\end{compactitem}

\descr{Methodology.} We start with a list of 122 cryptocurrency-related communities on Reddit, %
collecting 130.8M submissions and comments made on them. %
We then conduct a multi-axis analysis, performing temporal analysis and statistical modeling to analyze the evolution of the Reddit cryptocurrency communities over the years, both in terms of activity and user base.
We also study the evolution of discussions pertaining to specific cryptocurrencies while investigating the relationship between online activity and price movements.
Finally, using an emotion recognition classifier, we investigate the emotional dynamics of the cryptocurrency communities we study, 
and the relationship between emotions and market behavior.

\descr{Results.} Our main findings include:
\begin{compactenum}
    \item /r/CryptoCurrency and /r/dogecoin are the most active subreddits, featuring substantial and continuous posting activity, and a large number of active and new users. %
    Overall, there is a remarkable surge in cryptocurrency-related activity in 2021, characterized by heightened excitement and enthusiasm, followed by a sharp decline in interest by the end of the year, aligned with the rise and fall of several coins' value (\textbf{RQ1}).\myspace

    \item There is a strong relationship between activity and the price of Ethereum, Monero, Cardano, Polygon, and several other cryptocurrencies, with the changes in the number of posts mostly leading price changes (\textbf{RQ2}).\myspace

    \item \revision{Favorable real-world events like rising cryptocurrency values are correlated with joy, while a decline in the market is correlated with an increase in anger in user discussions.
    (\textbf{RQ3})}.\myspace
\end{compactenum}

Overall, our work presents a first-of-its-kind analysis of the ecosystem and highlights practical implications for market participants and regulators seeking to understand the evolving cryptocurrency landscape better.

\section{Related Work}
In this section, we review previous research focusing on i) characterizing cryptocurrency communities, %
ii) the interplay between price and online discussions,
and iii) emotional dynamics in the cryptocurrency ecosystem. %

Despite the significant growth of the cryptocurrency ecosystem, limited work has been done in the computational social science community to examine cryptocurrency-related communities on social media.
\citet{bohr2014uses} use publicly-available survey data of Bitcoin users to explore and understand the structure and characteristics of the Bitcoin community in its early stages.
\citet{glenski2019characterizing} analyze the growth and characteristics of online discussions surrounding Bitcoin, Ethereum, and Monero on Reddit from 2015 to 2017. 
\citet{linton2017dynamic} study  %
the interplay between opinions, topic evolution, and notable events on \url{bitcointalk.org}. 

Also,~\citet{johnson2023cryptocurrency} study the impact of cryptocurrency trading on mental health via qualitative analysis of /r/CryptoCurrency.
\citet{bonifazi2021social} use social network analysis to explore user behavior during a cryptocurrency speculative bubble.
Additional work uses cryptocurrency-related activity and discussion in social networks and instant-messaging platforms to forecast cryptocurrency prices or detect price manipulations~\cite{kim2016predicting,phillips2017predicting,glenski2019improved,hamrick2021examination}.
\revision{Others also utilize social media activity as a signal for predicting price movement in the stock market~\cite{xu2018stock}.}

\citet{wooley2019extracting} study 24 Reddit cryptocurrency communities to explore the connection between cryptocurrency prices and public sentiment.
They find that lagged price values and Reddit data can contribute to predicting Bitcoin and Ethereum price movements. %
\citet{telli2021multifractal} examine Bitcoin, Ethereum, Litecoin, and Ripple, investigating the multifractal characteristics and relationships between cryptocurrency price changes and user activity on Reddit and Wikipedia. 
Their findings reveal varying degrees of multifractality over different periods, predominantly anti-persistent behavior, and distinct dynamics in cross-correlations between social platforms and market factors.
\citet{lansiaux2022community} analyze the relationship between Twitter activity and changes in the price of Dogecoin and Litecoin.
\citet{stosic2018collective} focus on analyzing the cross-correlations between price changes of 119 publicly traded cryptocurrencies to measure collective behavior within the cryptocurrency market.
\revision{Finally,~\citet{phillips2018mutual} investigate relationships between Bitcoin and Ethereum price changes and topic discussions on Reddit, between August 2016 and August 2017, using dynamic topic modeling techniques and a Hawkes model.
They identify particular topics within these discussions that could serve as potential indicators of future price changes.}

\revision{Examining the relationship between markets and human emotions has been used in the context of financial decisions~\cite{chen2021influence}, predicting market behavior~\citet{wooley2019extracting}, and well-being~\cite{acheampong2020text}.
\citet{basile2021dramatic} explore how dramatic events in Reddit communities, like the COVID-19 pandemic, affect emotions, emphasizing the role of skepticism in danger mitigation.
In this context, our study also attempts to understand how cryptocurrency prices are correlated with users' emotions in cryptocurrency-related communities.}

\begin{table*}[t!]
\centering
\setlength{\tabcolsep}{4pt}
 \small
\begin{tabular}{l@{}rrrrr}
\toprule      
\textbf{Subreddit} &  \hspace{-0.4cm}\textbf{\#Submissions} &  \textbf{\#Comments} & \textbf{\#Tot. Posts} & \textbf{\#Users} & \textbf{Min. Date}\\ %
\midrule
 /r/CryptoCurrency &            1,574,333 &        30,799,627 &     32,373,960 &  844,878 & 2013-03-11\\% & 2022-08-31 \\
        /r/Bitcoin &            1,121,708 &        14,332,526 &     15,454,234 &  768,699 & 2010-09-10\\% & 2022-08-31 \\
       /r/dogecoin &            1,268,592 &        11,900,552 &     13,169,144 &  779,170 & 2013-12-08\\% & 2022-08-31 \\
      /r/ethtrader &              334,019 &         5,132,820 &      5,466,839 &  200,191 & 2015-03-25\\% & 2022-08-31 \\
/r/NFTsMarketplace &              343,555 &         5,076,311 &      5,419,866 &  152,571 & 2021-02-27\\% & 2022-08-31 \\
            /r/btc &              307,174 &         3,867,982 &      4,175,156 &  122,051 & 2012-06-16\\% & 2022-08-31 \\
       /r/SafeMoon &              273,500 &         3,434,480 &      3,707,980 &  122,978 & 2021-03-07\\% & 2022-08-31 \\
       /r/SHIBArmy &              271,048 &         2,453,562 &      2,724,610 &  138,577 & 2020-08-03\\% & 2022-08-31 \\
        /r/opensea &              164,677 &         2,499,077 &      2,663,754 &   97,307 & 2018-03-07\\% & 2022-08-31 \\
 /r/BitcoinMarkets &               50,033 &         2,445,312 &      2,495,345 &   67,177 & 2013-04-11\\% & 2022-08-31 \\
       /r/ethereum &              231,893 &         1,616,628 &      1,848,521 &  175,720 & 2014-01-05\\% & 2022-08-31 \\
       /r/Buttcoin &               93,428 &         1,713,617 &      1,807,045 &   63,739 & 2011-07-17\\% & 2022-08-31 \\
        /r/cardano &              104,817 &         1,229,902 &      1,334,719 &  108,627 & 2017-07-17\\% & 2022-08-31 \\
     /r/Crypto\_com &              110,673 &         1,216,529 &      1,327,202 &   74,736 & 2018-07-06\\% & 2022-08-31 \\
            /r/NFT &              328,942 &           969,900 &      1,298,842 &  150,135 & 2016-01-16\\% & 2022-08-31 \\
\bottomrule
\end{tabular}
\caption{Overview of the top 15 subreddits in our dataset. (Max date is 2022-08-31 for all.)} %
\label{tab:dataset_overview}
\end{table*}

Building on previous studies, we offer a more comprehensive understanding of cryptocurrency-related communities, by delving deeper in terms of both scale and timeframe.
\revision{
We strive to provide a comprehensive analysis of the cryptocurrency ecosystem by studying diverse cryptocurrency communities and investigating 50 cryptocurrencies, including well-known ones like Bitcoin and Ethereum, as well as more questionable ones like Terra LUNA. 
This approach broadens our understanding of the ecosystem's evolution, and the interplay between certain events, social media posting activity and emotions, and price movements.
In contrast to prior research, which examines only a portion of cryptocurrency lifetimes, we analyze the entire lifespan of our selected cryptocurrencies until August 2022. 
This period encompasses significant price fluctuations, allowing us to uncover various links between price changes, social media activity, and user emotions.
Additionally, while others focus on general user sentiment and its relation to cryptocurrency prices, we narrow our focus to specific emotions. Our aim is to explore how cryptocurrency events and price shifts align with the emotional tenor of discussions within cryptocurrency communities.
}
Overall, our analysis is the first to shed light on the intricate connections between user engagement, emotional responses, and market behavior, thus contributing to a more nuanced comprehension of the cryptocurrency ecosystem.

\section{Dataset}

In this section, we present the data collection and annotation process to collect posts published on cryptocurrency-related communities on Reddit.

\subsection{Identifying Cryptocurrency Communities}
Our first step is to identify a set of cryptocurrency-related communities, or {\em subreddits}, on Reddit. 
We then get all submissions and comments from these subreddits using Pushshift's monthly dumps~\cite{baumgartner2020pushshift}.
We focus on Reddit given its popularity within the cryptocurrency community~\cite{cryptocurrency2021reddit}.

We start from the top 50 cryptocurrencies for market capitalization, obtained from CoinMarketCap.com on March 13, 2022.\footnote{The full list is available from \url{https://drive.google.com/file/d/1ynkoO9muO7AS72mOzsdycRYAIKvy6Coh/view}} %
CoinMarketCap is a platform that provides cryptocurrency market cap rankings and real-time tracking of cryptocurrency prices.
We consider the top 50 cryptocurrencies as their marketcap distribution constitutes the majority of all the cryptocurrencies available up to March 13, 2022. %

We collect all comments made on Reddit between September 2021 and May 2022 that mention at least one of the top 50 cryptocurrency names.
This yields a total of 2.96M comments posted in 49.7K unique subreddits.

\descr{Subreddits Annotation.}
Next, we sort the 49.7K subreddits based on the number of comments mentioning a cryptocurrency normalized by the total number of comments posted in these subreddits during the same period and focus on the top 1K for manual annotation.
Each of the top 1K subreddits is presented to three annotators, along with the cryptocurrency mentioned the most in the comments.
The annotators inspect the subreddit page, the community's description, as well as the submissions and comments, and label each subreddit as:

\begin{compactitem}
\item \textbf{\em Relevant,} if its description and discussions on the subreddit are primarily about a specific cryptocurrency (e.g., /r/bitcoin) or about the cryptocurrency market in general (e.g., /r/CryptoCurrency).
We also consider as relevant any subreddit related to other cryptocurrencies, even if not among the top 50 cryptocurrencies for market cap; or
\item \textbf{\em Irrelevant, } if its description and posts contain content unrelated to cryptocurrencies.
For example, subreddits that are primarily focused on science, technology, music, as well as stocks or trading are considered irrelevant, even if some discussion mentions some cryptocurrencies.
\end{compactitem}

\descr{Annotation Details.}
\revision{The manual annotation process is carried out by three non-student authors of this paper, all familiar with cryptocurrencies and with previous research experience in the field.}
Each annotator labels each subreddit, and we go with the majority agreement.
This results in {\em 122 relevant and 878 irrelevant subreddits}.

The annotation yields a Fleiss' Kappa score ($k$)~\cite{fleiss1971measuring}, which assesses the agreement between the annotators, of $k=0.96$; this is considered a ``{\em very good}'' agreement~\cite{landis1977measurement}.

\begin{figure*}[t!]
\centering
 \begin{subfigure}[b]{0.49\textwidth}
    \includegraphics[width=\linewidth]{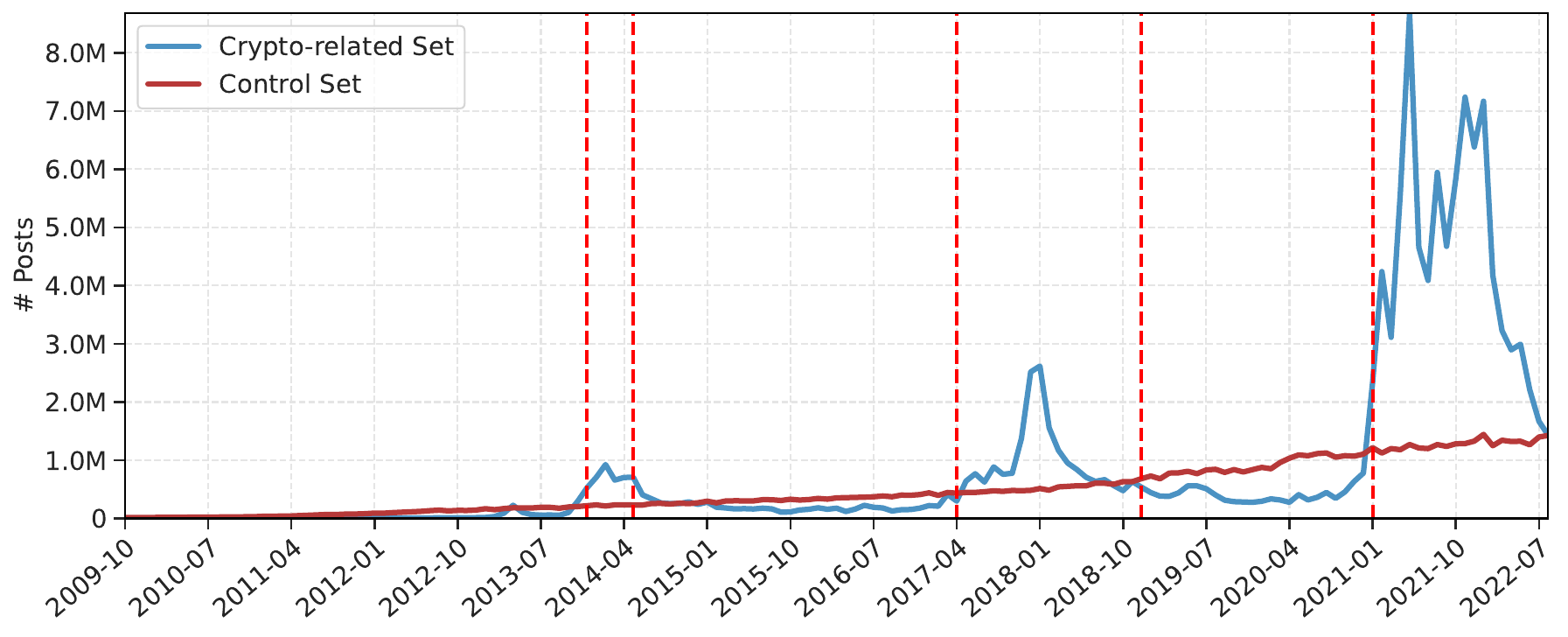}
    \caption{absolute number of posts (submissions+comments)}\myspace
 \end{subfigure}
 ~
\begin{subfigure}[b]{0.48\textwidth}
    \includegraphics[width=\linewidth]{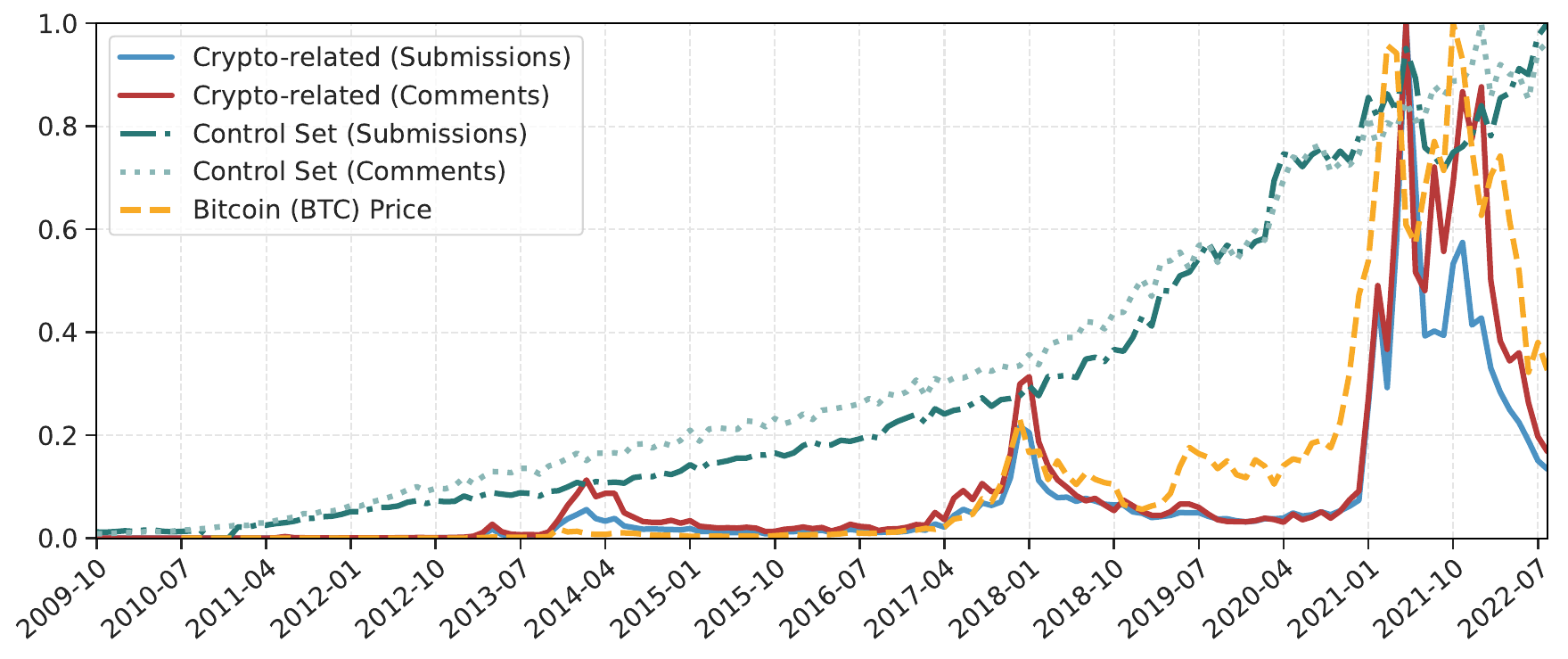}
    \caption{normalized submissions and comments and Bitcoin price}\myspace
\end{subfigure}
\myspace\caption{Temporal evolution of the number of posts (monthly) in our dataset.} 
\label{fig:posts_temporal_evolution}
\end{figure*}

\begin{figure*}[t!]
\centering
\begin{subfigure}[b]{0.49\textwidth}
	\includegraphics[width=\linewidth]{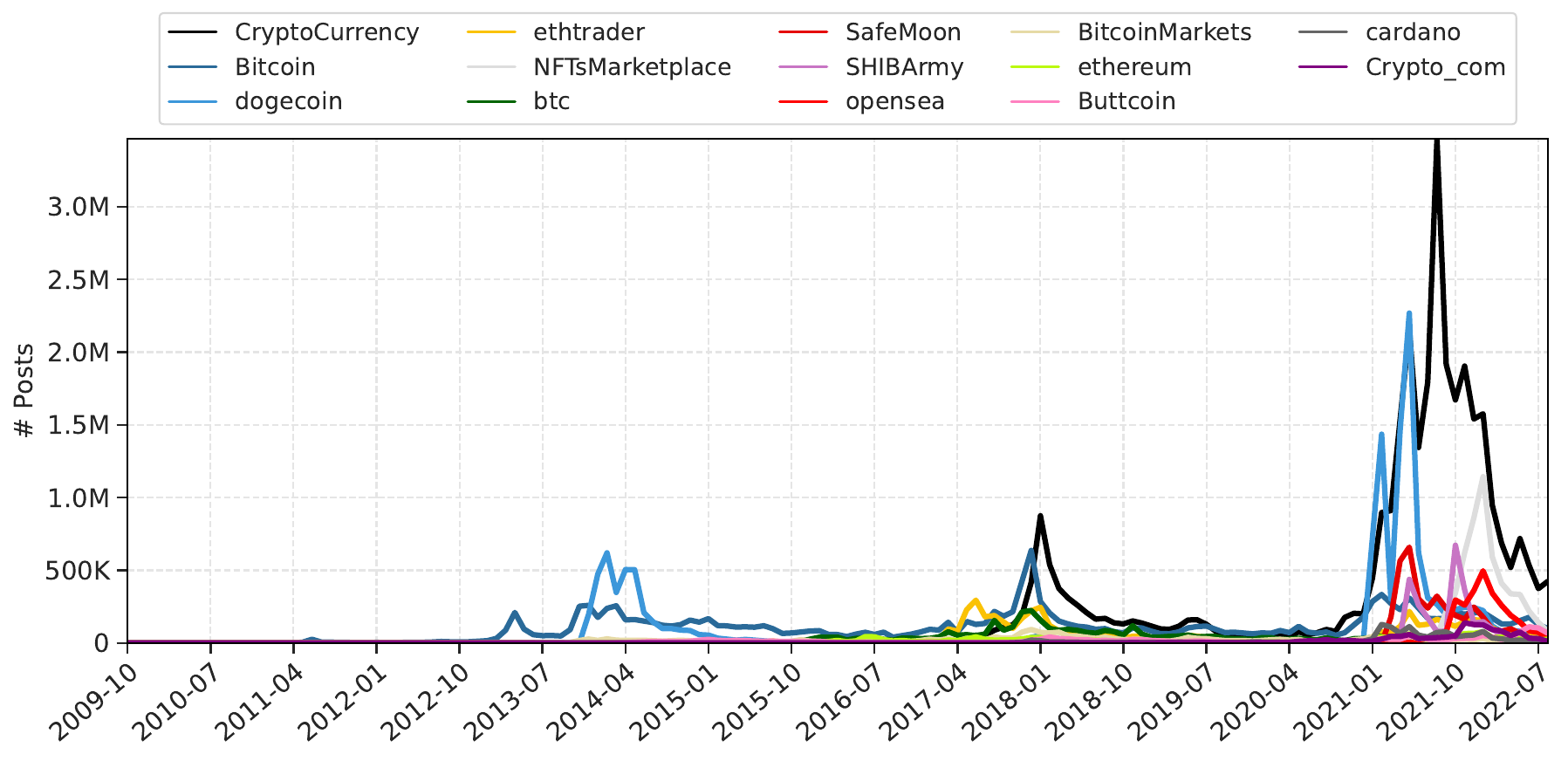}
	\caption{absolute}\myspace 
\end{subfigure}	
~
\begin{subfigure}[b]{0.48\textwidth}
	\includegraphics[width=\linewidth]{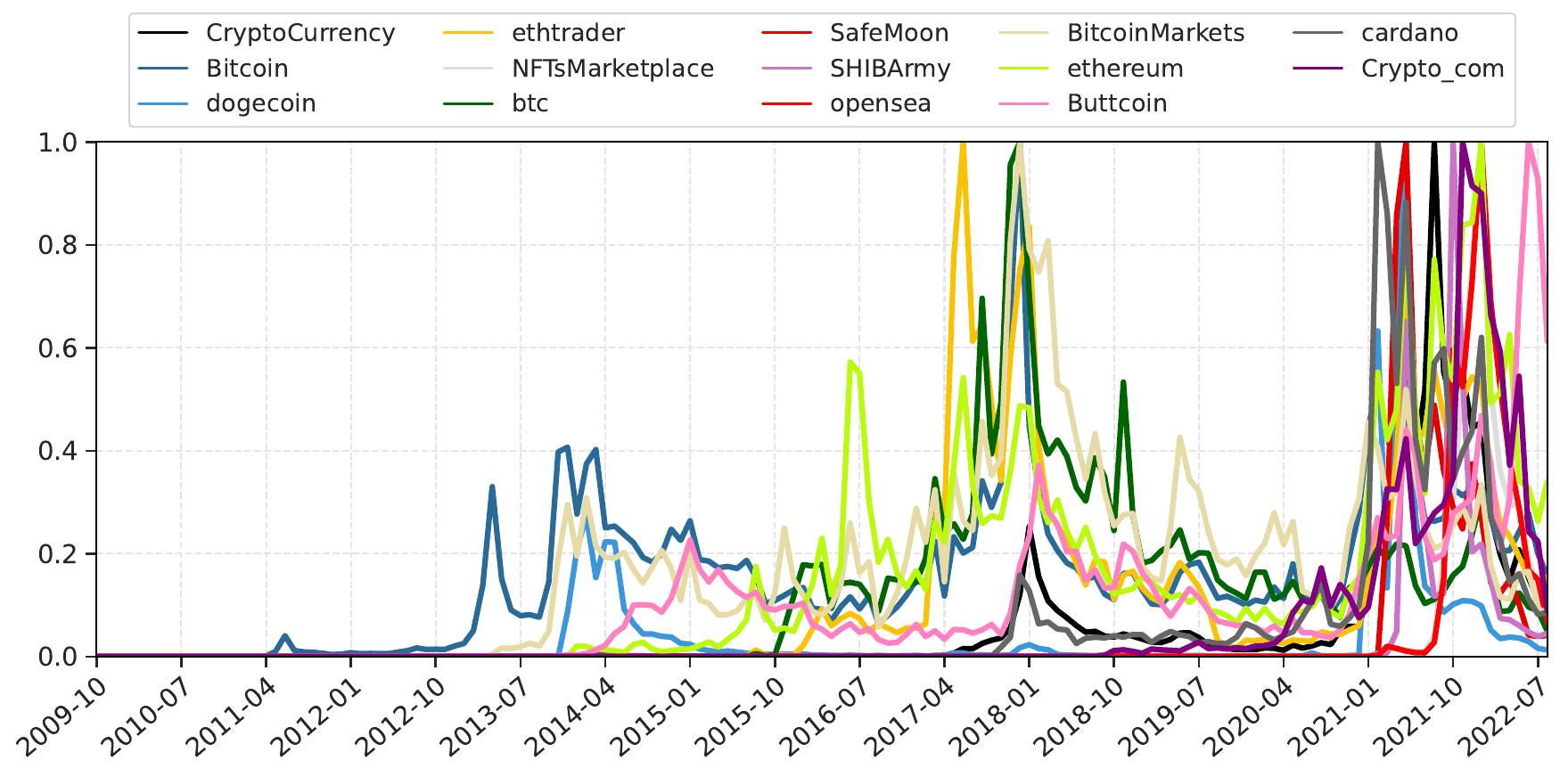}
	\caption{normalized}\myspace 	
\end{subfigure}	
\myspace\caption{Temporal evolution of the monthly number of posts for the top 15 subreddits in our cryptocurrency-related dataset.}
\label{fig:posts_temporal_evolution_subreddits}
\end{figure*}

\subsection{Data Collection}
\label{subsec:data_collection}

Next, we collect all submissions and comments made between June 1, 2005 and August 31, 2022 from the 122 subreddits identified as relevant to cryptocurrencies, using Pushshift's Reddit monthly dumps~\cite{baumgartner2020pushshift}. 
Note that we collect all submissions and comments at this step, not just those mentioning a cryptocurrency.

This yields a total of 11.5M submissions and 119.3M comments.
Throughout the rest of this paper, we refer to this set of submissions and comments as the \textit{\textbf{``cryptocurrency-related''}} set of posts.
Table~\ref{tab:dataset_overview} reports the total number of posts (submissions and comments), submissions, comments, users, and the period of available information; to ease presentation, the table only reports the top 15 subreddits, while the complete list is available from~\cite{anonymous2023dataset}.

The first cryptocurrency-related subreddit in our dataset ever created is /r/Ripple (top 25 subreddit and top 6 coin as per market cap), on October 14, 2009.
In general, /r/CryptoCurrency is the most popular cryptocurrency-related subreddit, with more than double the number of posts of the second subreddit in our dataset, /r/Bitcoin.
Interestingly, the third most popular subreddit in terms of posts is /r/dogecoin, a community dedicated to the ``memecoin'' Dogecoin, which also has the second largest user base.

\descr{Control Set.}
We collect and use as a control set, a random sample from Reddit, consisting of 0.5\% of the entire Reddit corpus (between June 1, 2005 and August 31, 2022), which amounts to 75.6M posts.

\descr{Historical Price Data.}
We obtain the historical daily price data for the 50 cryptocurrencies we consider, from their initial listing date until August 31, 2022, from \href{https://eodhistoricaldata.com/}{https://eodhistoricaldata.com/}.
More precisely, we only consider each cryptocurrency's daily \textit{close} price.

\descr{Ethics Considerations.} Since we do not work with human subjects and only use data available to the public, our work is not categorized as human subjects research by our institution's Institutional Review Board (IRB).
Overall, we follow standard ethical guidelines regarding information research and the use of shared measurement data. 
\section{Longitudinal User-Level Analysis}
In this section, we discuss our user-level analysis of the Reddit cryptocurrency communities.

\begin{figure*}[t!]
\centering
\begin{subfigure}[b]{0.49\textwidth}
    \includegraphics[width=\linewidth]{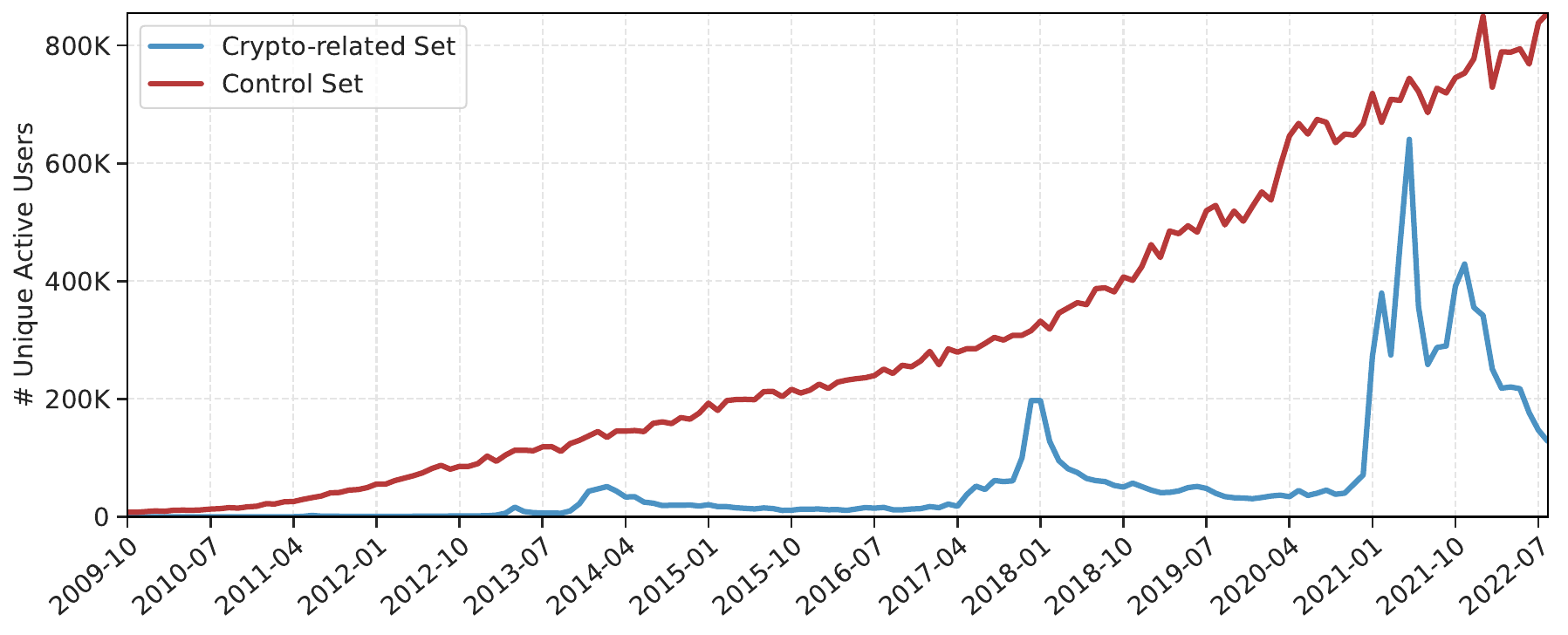}
    \caption{absolute}\myspace\myspace   
\end{subfigure}
~
\begin{subfigure}[b]{0.48\textwidth}
    \includegraphics[width=\linewidth]{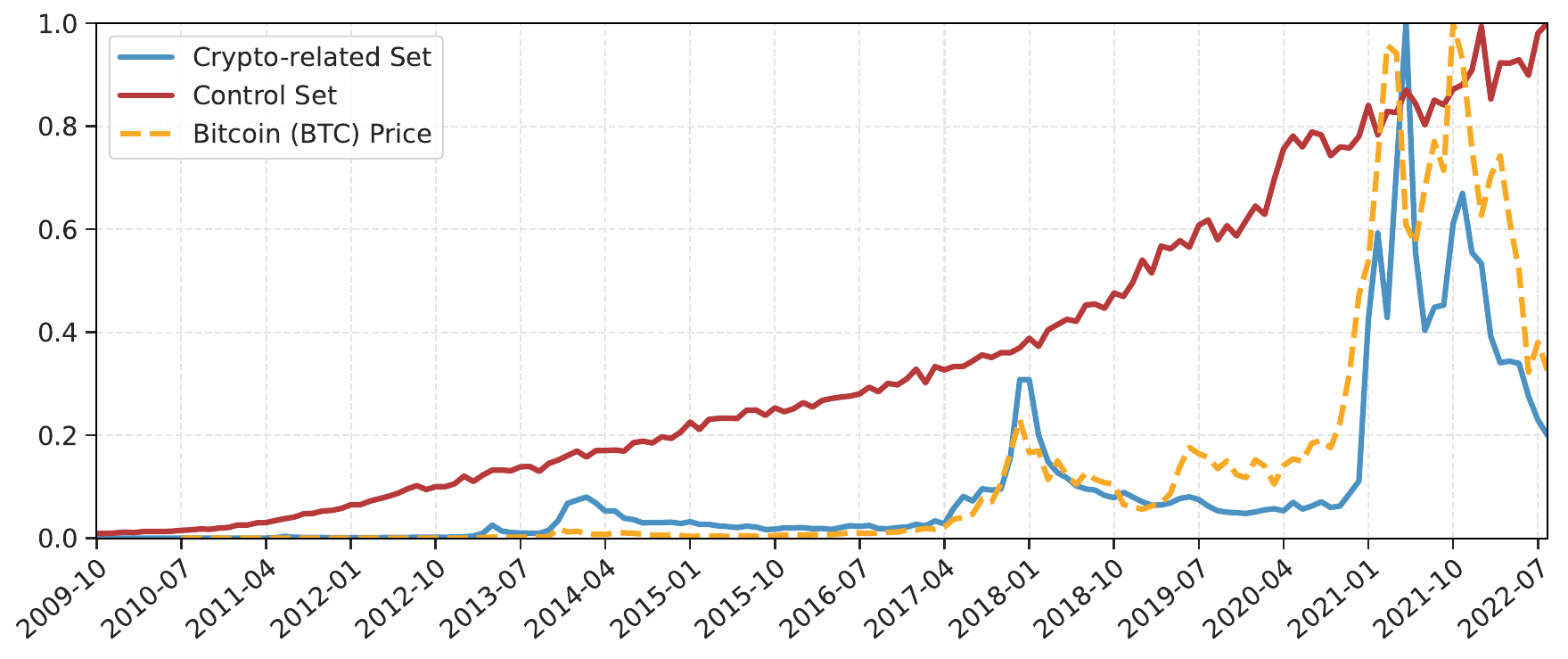}
    \caption{normalized}\myspace\myspace 
\end{subfigure} \\
\begin{subfigure}[b]{0.49\textwidth}
    \includegraphics[width=\linewidth]{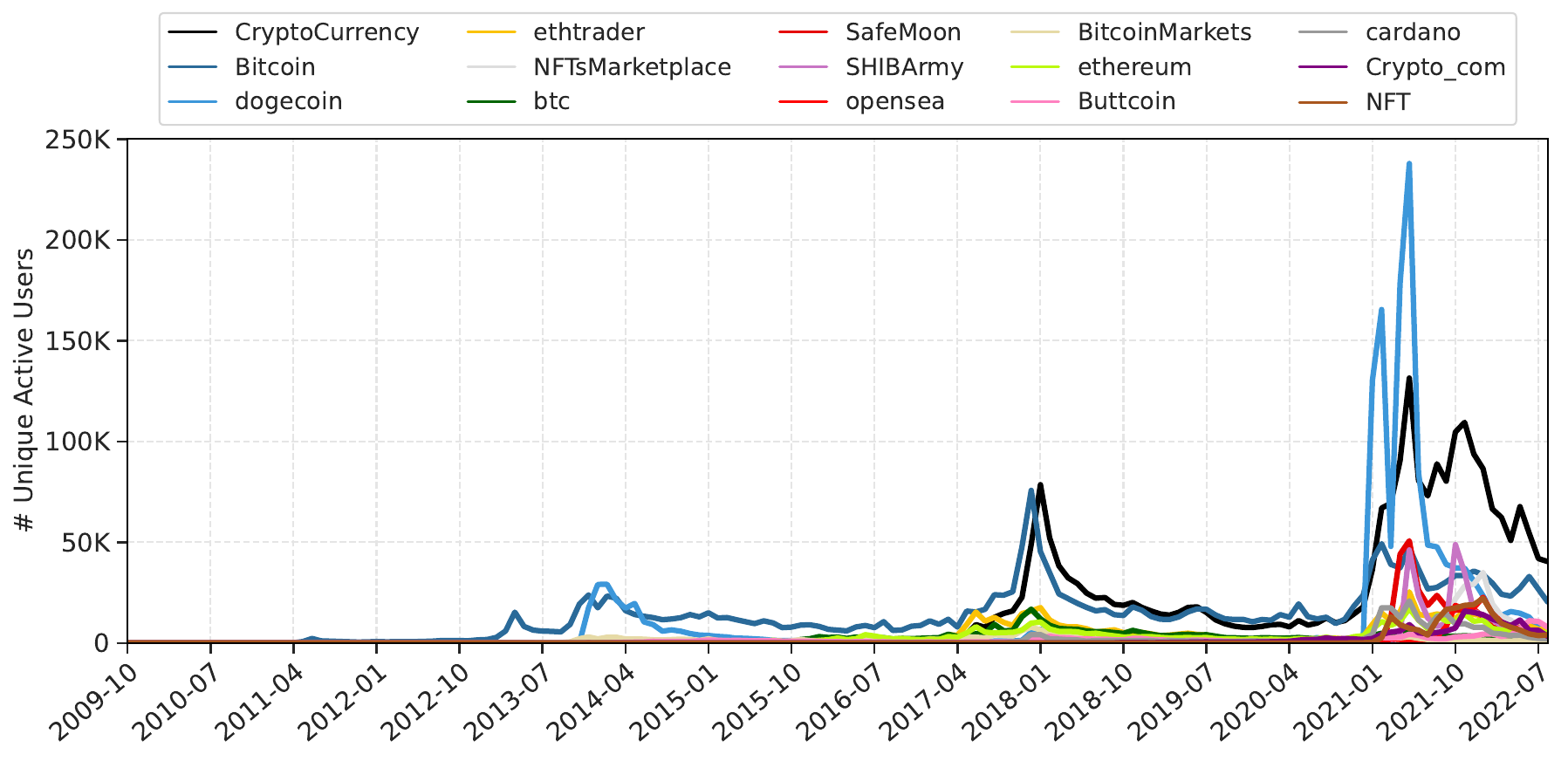}
    \caption{absolute per subreddit}\myspace\myspace 
\end{subfigure}
\begin{subfigure}[b]{0.48\textwidth}
    \includegraphics[width=1\linewidth]{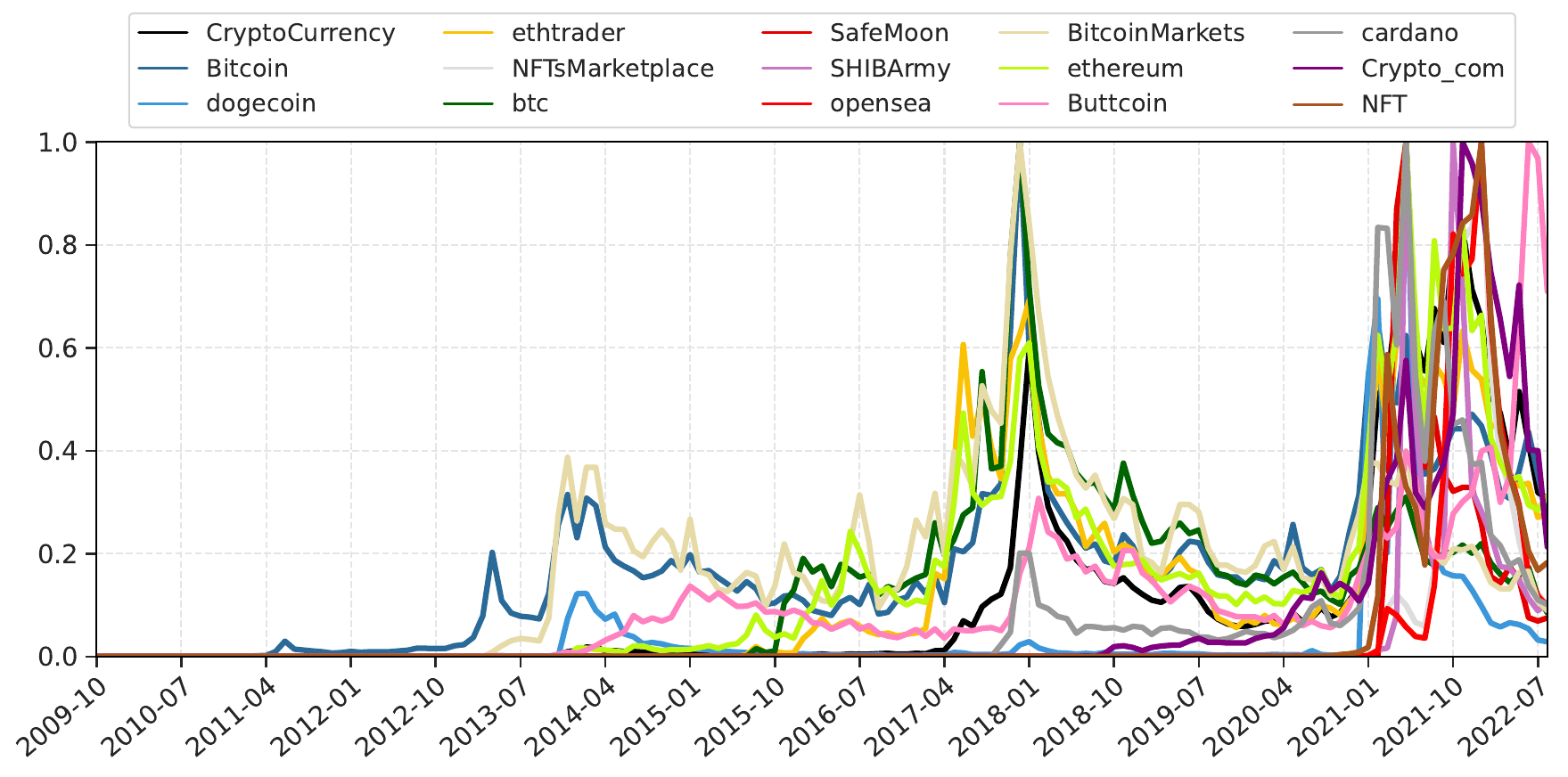}
    \caption{normalized per subreddit}\myspace\myspace 
\end{subfigure} 
\caption{Evolution of the number of monthly unique active users in our dataset.}%
\label{fig:unique_users_temporal_evolution}
\end{figure*}

\subsection{User Activity}
We start with a longitudinal analysis of the number of posts made on cryptocurrency-related subreddits. 
Figure~\ref{fig:posts_temporal_evolution} plots the evolution of the number of posts over time, as well as the normalized price of Bitcoin.
Red vertical dotted lines indicate events identified by changepoint analysis.
\revision{Changepoint detection is a statistical technique used to identify points or time periods where a significant change or discontinuity has occurred in a given time series.
In this study, it allows us to detect and investigate the possible reasons behind significant changes in user activity or in a cryptocurrency's market price.} 
Normalization is done by dividing each time series by its maximum, aiming to retain the original relationships between the data points while comparing unrelated sequences.
We also plot the normalized Bitcoin price to uncover any relationship between the changes in the posting behavior and the price of Bitcoin, performing changepoint detection on the posting activity of the cryptocurrency-related set using the PELT algorithm~\cite{killick2012optimal}.
Overall, there is an average of 74.1K submissions and 769.4K comments being made every month in these communities.

\descr{Peaks.} We observe a first peak in the total activity between December 2013 and May 2014, around the bankruptcy of the MtGox platform~\cite{mtgoxbankruptcy2014reuters} with 226K posts just in March 2014.
The second important peak occurs between April 2017 and December 2018, with 2.6M posts just in December 2017, around the launch of the Bitcoin futures contract and the first historical price record.
The last peak period happens in 2021 with: i) Bitcoin's market capitalization passing $\$1T$ in February 2021~\cite{bitcoin2021cnbc}, ii) interest in and price surge of Dogecoin after Elon Musk's tweets in February 2021~\cite{dogecoin2021elon}, and iii) increased interest in non-fungible tokens (NFTs) after Beeple's \$69M sale in March 2021~\cite{beeplenft2021cnbc}.
Just in the month of April 2021, we count a total of 9.8M posts.

In general, after the end of 2013 (the first listing price of Bitcoin), the monthly number of posts in our cryptocurrency-related set follows similar trends as the Bitcoin price compared to the smoother increase in the number of posts in the control set over time.

\descr{Top Coins.} Figure~\ref{fig:posts_temporal_evolution_subreddits} presents the evolution of the absolute and normalized number of posts for the top 15 cryptocurrency-related subreddits in our dataset.
As expected, /r/CryptoCurrency has the most activity since it is also the biggest subreddit and the one driving the peak in activity between 2021 and 2022.
Notably, there is also a sudden peak in /r/dogecoin's activity in February 2021 (around Dogecoin's 216\% price increase after Elon Musk's Dogecoin-related tweets); interestingly, the peak in /r/dogecoin's activity happens \emph{before} the price increase.

\subsection{Active Users} 

\descr{Removing Spam Users.} 
To eliminate spam and capture more organic popularity trends, we remove all posts made by the ``Automoderator'' user, a bot used by Reddit creators to automatically moderate content on the platform.
We also use the `spam' score from Perspective API~\cite{perspective2023}; we consider a post to be spam if the score is above 0.9; then, we calculate the percentage of spam posts per (unique) user.
To find an appropriate threshold to exclude spammy users from our analysis, we start by excluding posts from users with over $90\%$ spam posts and have two of the authors annotate 100 randomly selected posts from the remaining ones -- only if both agree that a post is spam, we consider it spam. 
\revision{More precisely, we consider a post to be spam if it appears to be posted by a bot.}

We iteratively repeat this process excluding posts from users with a decreasing percentage of spam posts, down to $10\%$, and calculate the percentage of posts annotated as spam in each set.
After excluding posts from users with over $50\%$ spam posts, the percentage of spam in the remaining dataset drops a lot ($\le 8\%$); as a result, we exclude all users with over $50\%$ spam posts.
More precisely, we exclude posts from 200K users ($5.5\%$) from the 3.6M unique users in our cryptocurrency-related dataset.
\revision{
We do so to discern and exclude all automated bot users from our user-level analysis and thus better capture the genuine trends in popularity driven by (human) users.
}

\begin{figure*}[t!]
\centering
\begin{subfigure}[b]{0.49\textwidth}
    \includegraphics[width=\linewidth]{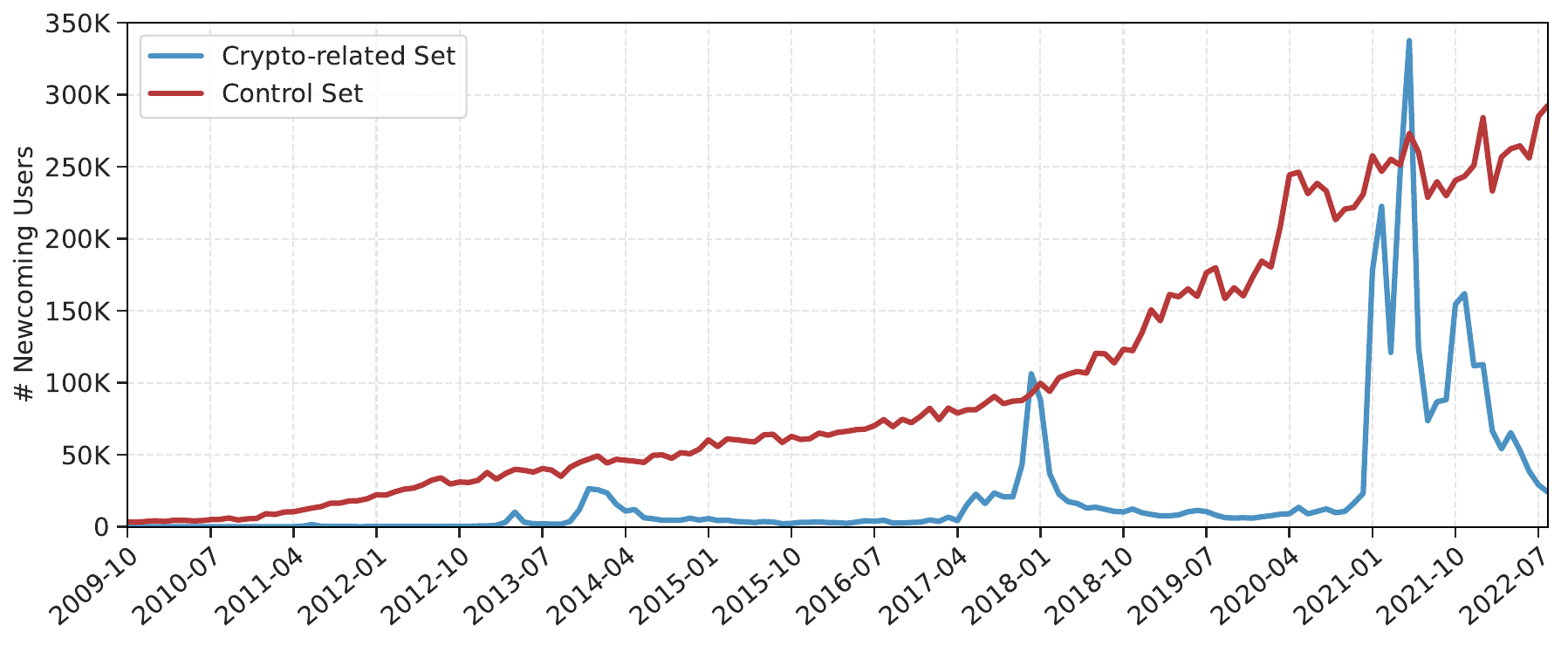}
    \caption{absolute}\myspace\myspace    
\end{subfigure}
~
\begin{subfigure}[b]{0.48\textwidth}
    \includegraphics[width=\linewidth]{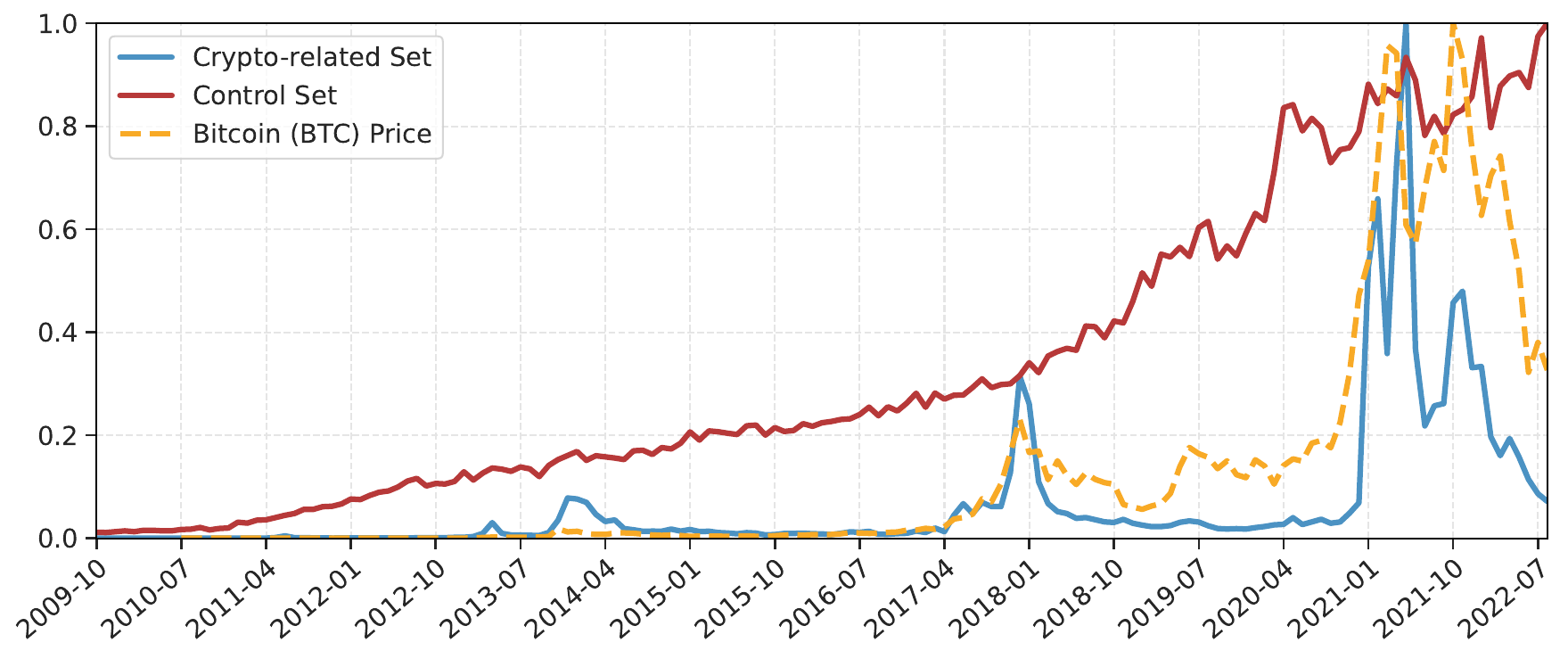}
    \caption{normalized}\myspace\myspace     
\end{subfigure}
\begin{subfigure}[b]{0.49\textwidth}
    \includegraphics[width=\linewidth]{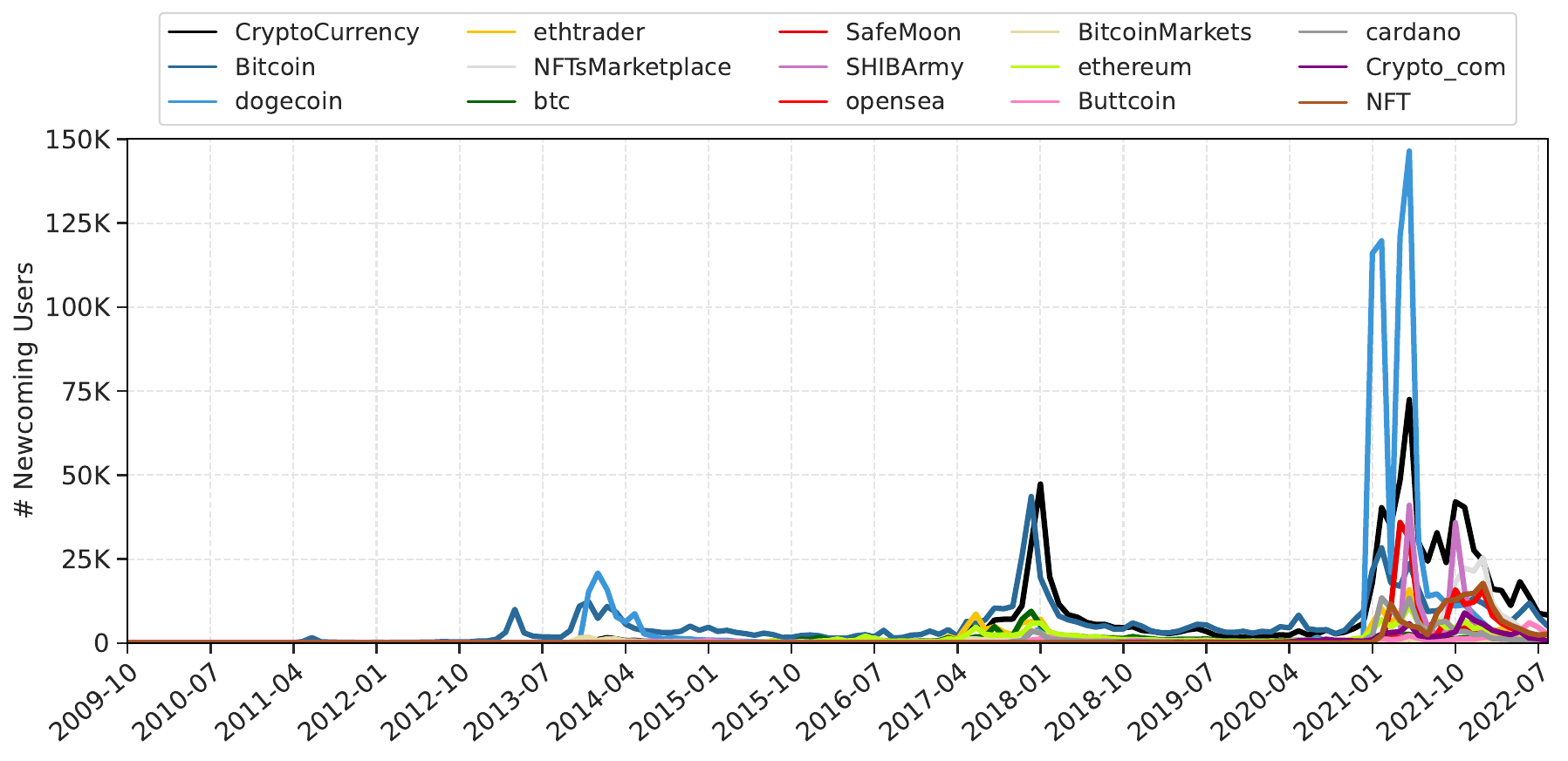}
    \caption{absolute per subreddit}\myspace\myspace     
\end{subfigure}
~
\begin{subfigure}[b]{0.48\textwidth}
    \includegraphics[width=\linewidth]{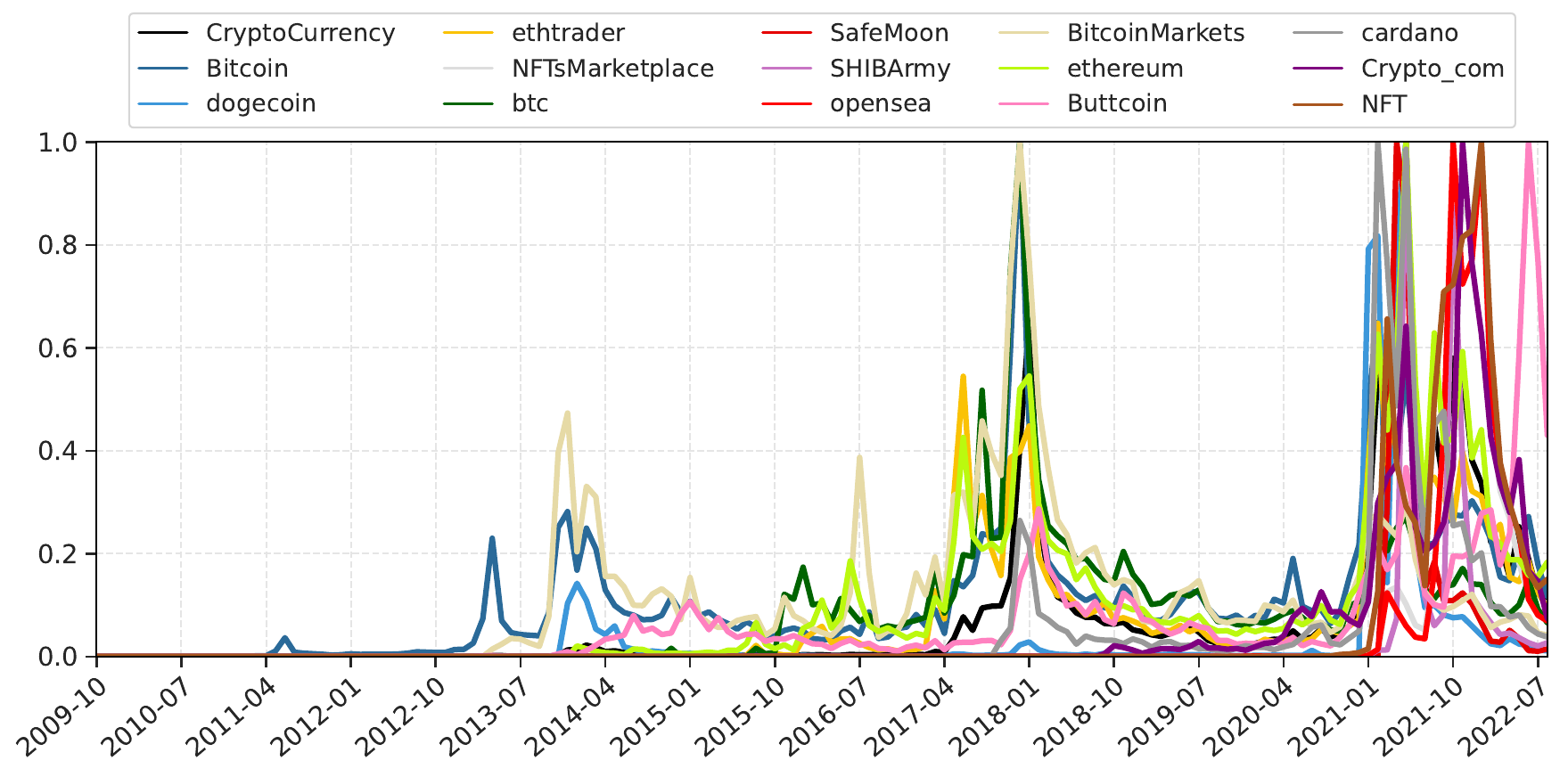}
    \caption{normalized per subreddit}
\end{subfigure} 
\caption{Evolution of the number of monthly new users in our dataset.}
\label{fig:newcoming_users_temporal_evolution}
\end{figure*}

\descr{Monthly Active Users.} %
We consider users to be active in a given month if they made at least one submission or comment.
Figure~\ref{fig:unique_users_temporal_evolution} reports the absolute and normalized number of unique monthly active users in the cryptocurrency-related and control sets and the top 15 subreddits.
While the number of unique active users rises steadily and smoothly on Reddit (as per our control set), this is not the case for the cryptocurrency-related subreddits.
Instead, we observe an initial increase at the end of 2017 and a second increase after the end of 2020, but overall the number fluctuates.

When focusing on the top 15 subreddits, we find that /r/CryptoCurrency, /r/Bitcoin, and /r/dogecoin have the largest user base.
Also, /r/bitcoin has most of its active users in December 2017, when the price of Bitcoin hit a new all-time high of \$20K.
The number of active users drops around March 2018, as the Bitcoin price also drops.
For /r/CryptoCurrency and /r/dogecoin, there is a sharp increase in active users during the first half of 2021 when Bitcoin reached its current all-time high of \$67.6K.
For /r/dogecoin, there are two sharp increases during this period, one in February 2021 (165K active users), which is probably associated with Elon Musk's Dogecoin-related tweets, and another one in May 2021 (238K active users) when Dogecoin reached its all-time high price of \$0.74.

\subsection{New Users}
Next, we examine the number of {\em new} monthly users; we consider users to be ``new'' at any given point in time if they have never made a submission or a comment previously.
In Figure~\ref{fig:newcoming_users_temporal_evolution}, we report the temporal evolution of the absolute and normalized number of new users. %
In general, we observe similar trends with unique active users.
However, here /r/dogecoin has the highest number of new users.
More precisely, /r/dogecoin exhibits a sharp increase in new users in January 2021 (15,857\%) and April 2021 ($474\%$).
Incidentally, the former coincides with Elon Musk's Dogecoin tweets~\cite{dogecoin2021elon}, the latter with a surge in Dogecoin's price ahead of Elon Musk's appearance on ``Saturday Night Live'' in May 2021~\cite{muskdogesnl2021cnbc}.

\descr{User Retention.}
To examine whether the sharp increase in unique active and new users in the cryptocurrency-related subreddits after 2020 is due to the increased interest by general users (i.e., the cryptocurrency community going more ``mainstream'') or to an increased interest in those communities and their discussions by the {\em same} users, 
we measure user retention over time using the Szymkiewicz-Simpson coefficient~\cite{vijaymeena2016survey}.

\begin{figure}[t!]
\centering
\includegraphics[width=0.99\columnwidth]{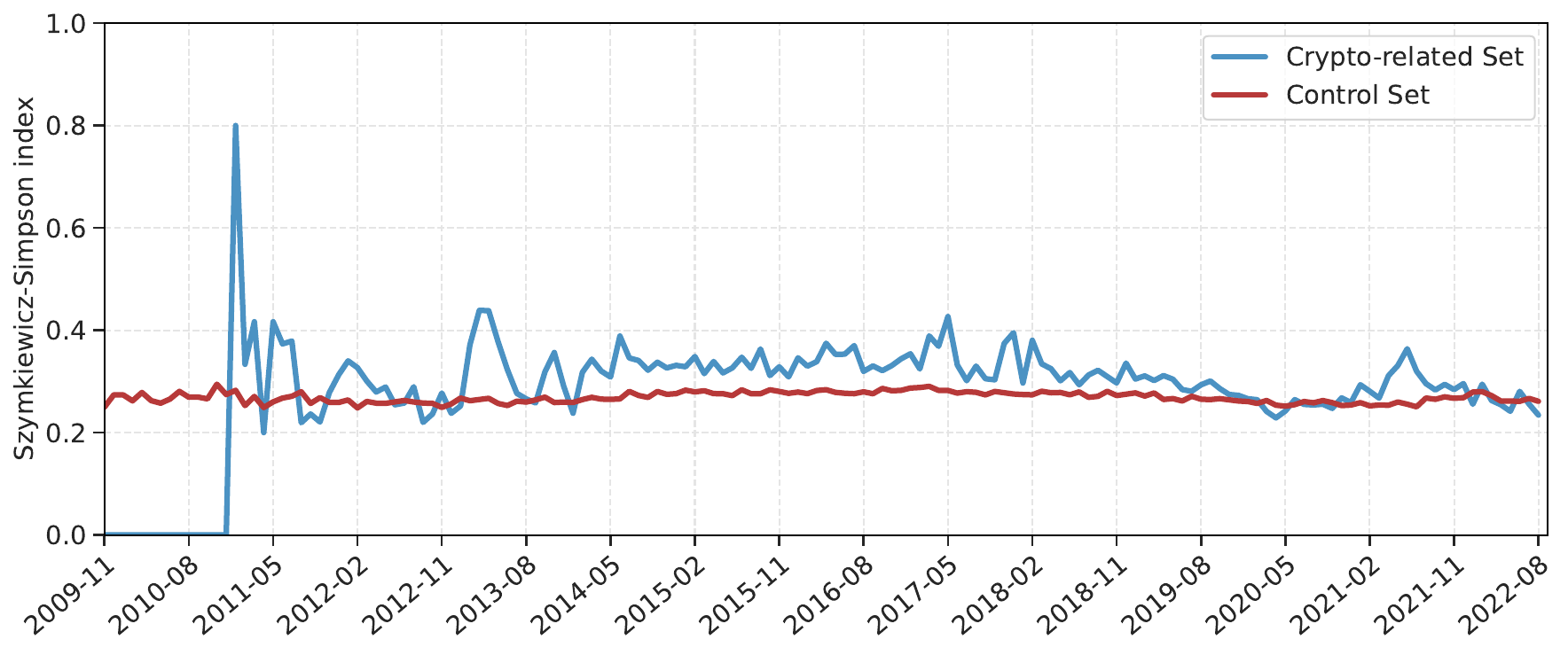}
\caption{Self-similarity of commenting users in adjacent months.} %
\label{fig:users_overlap}
\myspace
\end{figure}

Specifically, we measure the similarity between posting users in a given month and those doing so the \textit{month before}.
If the set of commenting users in a specific month is a subset of the previous month's commenting users (or the converse), the overlap coefficient equals 1.
The overlap coefficients are plotted in Figure~\ref{fig:users_overlap}.
While user retention remains the same over time for the control set, this is not the case for the user base of the cryptocurrency-related set.
More precisely, the latter exhibits several peaks and drops over time, especially after 2021, which probably indicates the increased interest in cryptocurrencies by random users and that cryptocurrencies are becoming more mainstream over time.

\begin{figure*}[t!]
\centering
\begin{subfigure}[b]{0.49\textwidth}
    \includegraphics[width=\linewidth]{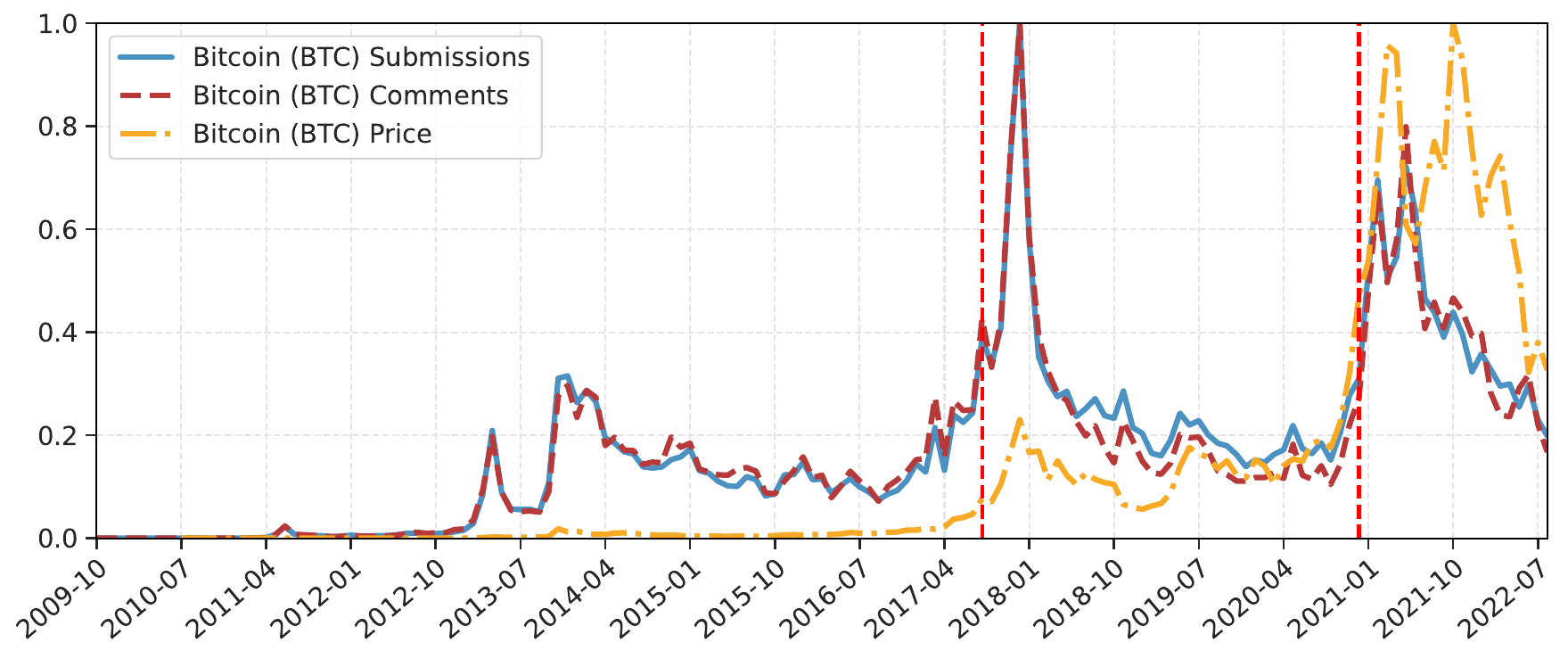}
    \caption{Bitcoin (BTC)}\myspace\myspace
\end{subfigure}
\begin{subfigure}[b]{0.49\textwidth}
    \includegraphics[width=\linewidth]{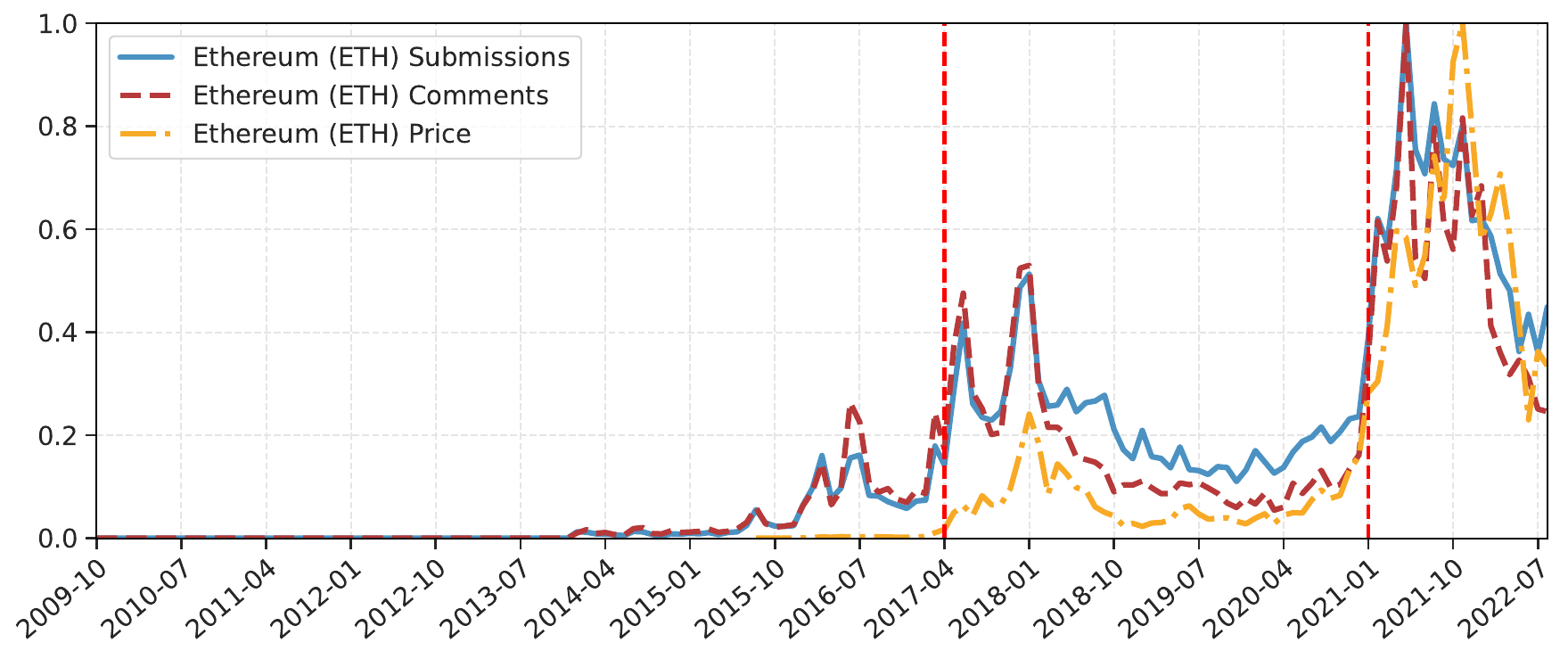}
    \caption{Ethereum (ETH)}\myspace\myspace
\end{subfigure}
\begin{subfigure}[b]{0.49\textwidth}
    \includegraphics[width=\linewidth]{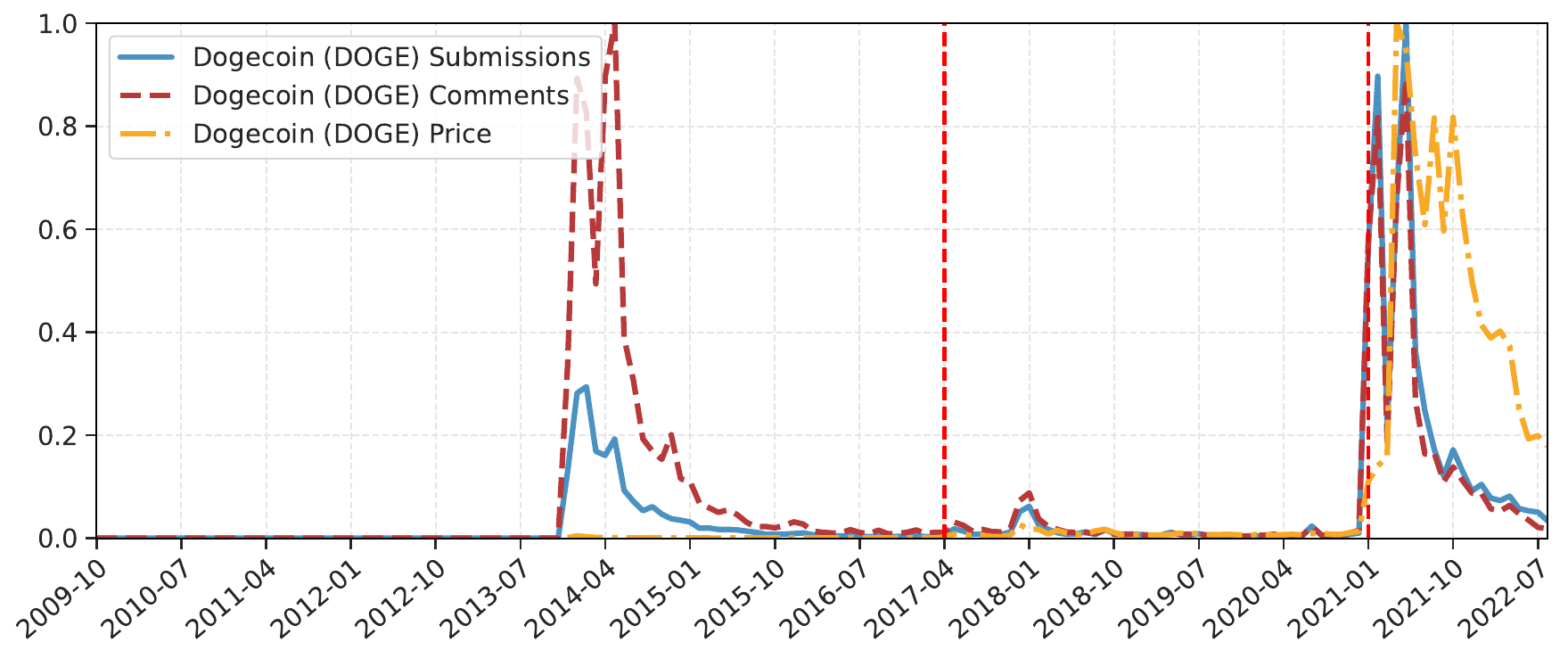}
    \caption{Dogecoin (DOGE)}\myspace\myspace
\end{subfigure}
\begin{subfigure}[b]{0.49\textwidth}
    \includegraphics[width=\linewidth]{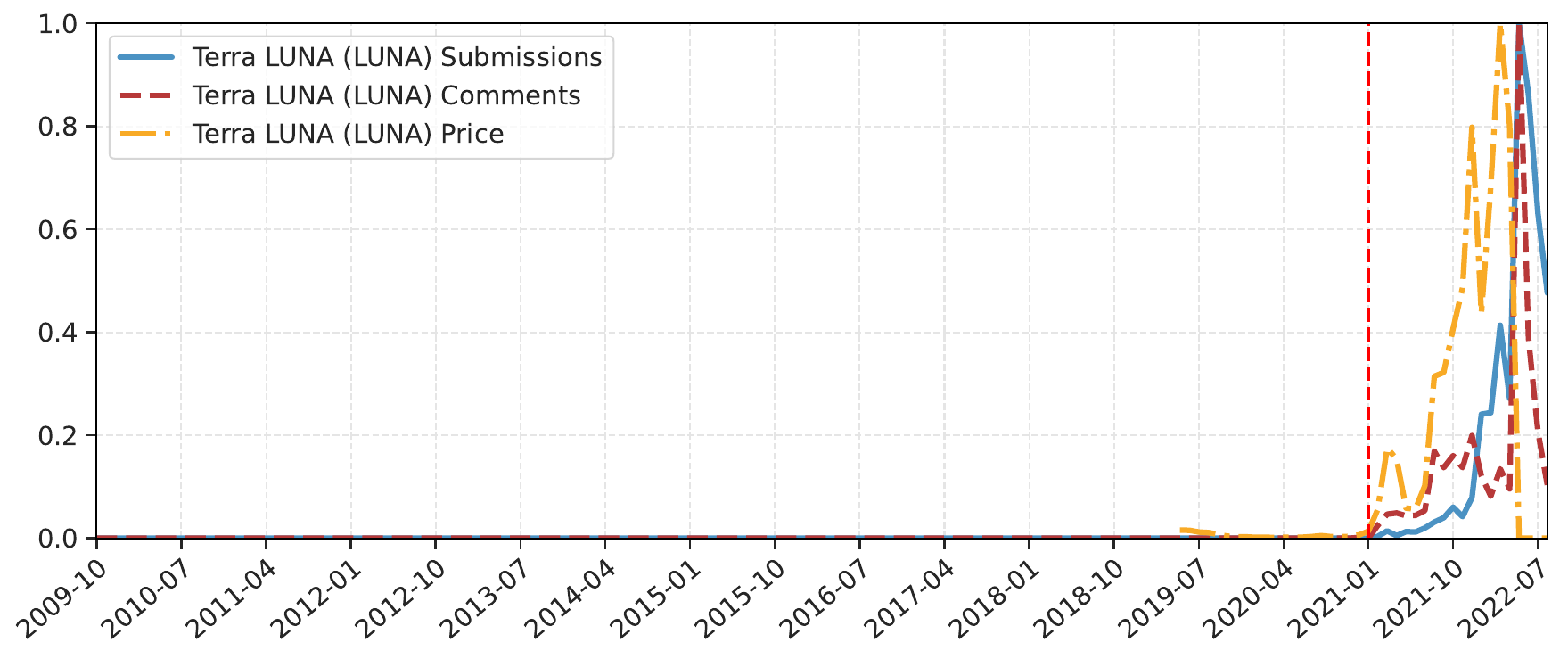}
    \caption{Terra LUNA (LUNA)}\myspace\myspace
\end{subfigure}
\begin{subfigure}[b]{0.49\textwidth}
    \includegraphics[width=\linewidth]{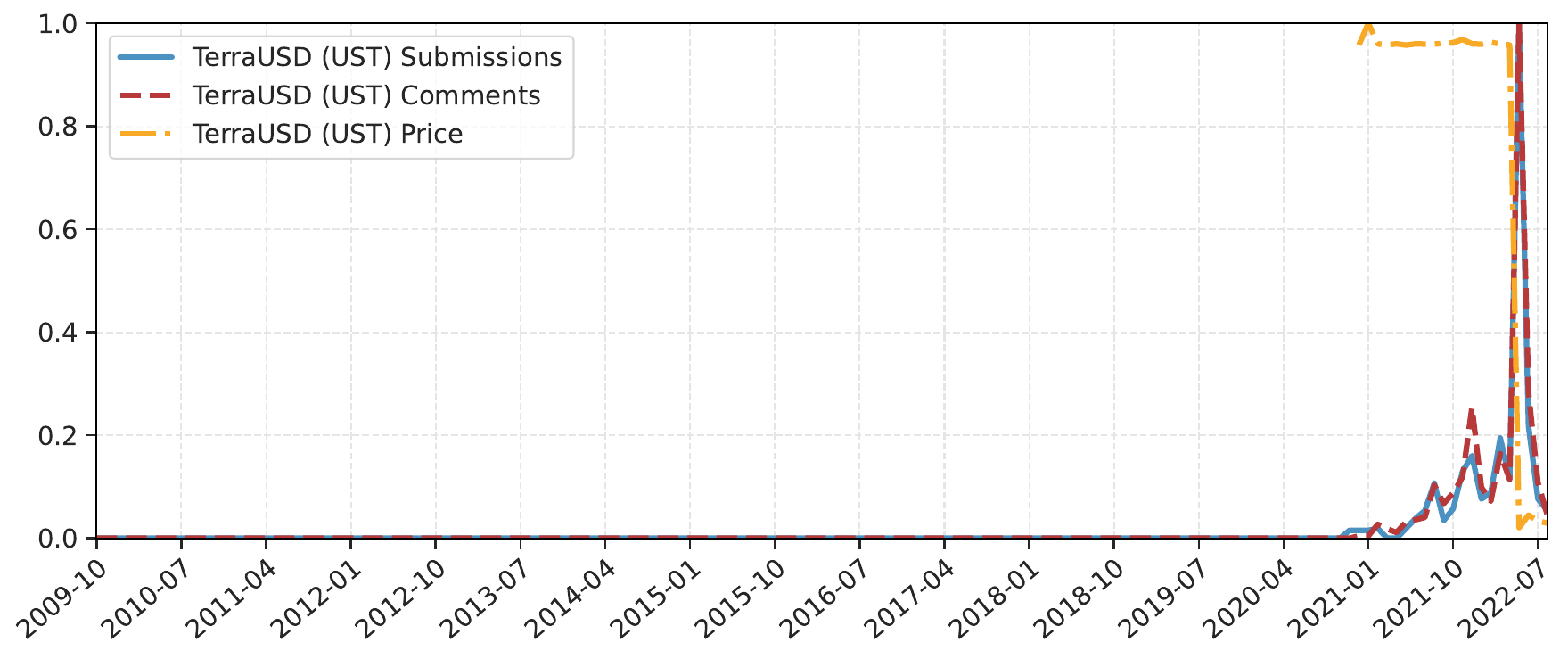}
    \caption{TerraUSD (UST)}\myspace\myspace
\end{subfigure}
\begin{subfigure}[b]{0.49\textwidth}
    \includegraphics[width=\linewidth]{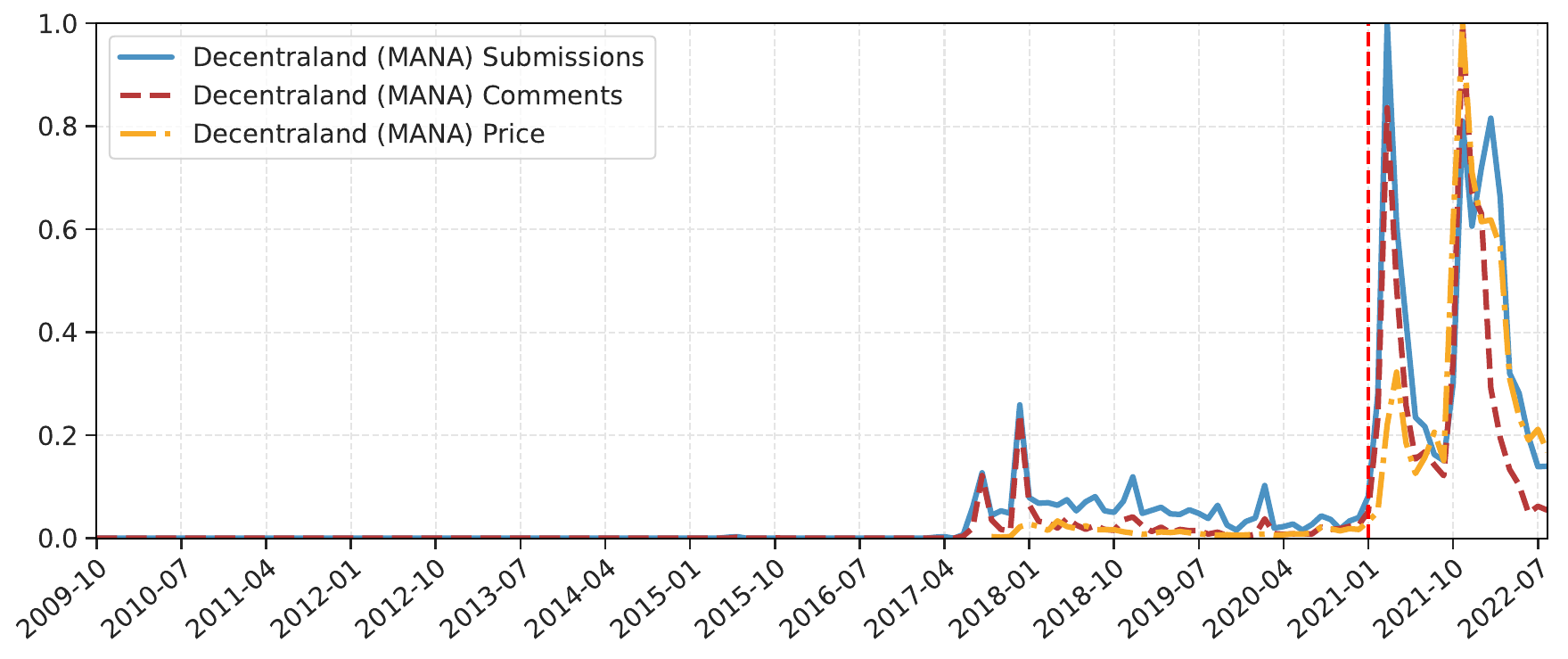}
    \caption{Decentraland (MANA)}\myspace\myspace
\end{subfigure}    
\caption{Temporal evolution of the monthly normalized number of submissions and comments mentioning cryptocurrencies in our dataset along with the price of the corresponding cryptocurrency.}
\label{fig:crypto_monthly_posts}
\end{figure*}

\begin{table}[t]
\centering
\small
\setlength{\tabcolsep}{3pt}
\begin{tabular}{l@{}rrrrr}
\toprule
\textbf{Currency} & \hspace{-0.25cm}\textbf{\#Submissions} &  \textbf{\#Comments} & \textbf{\#Posts} &   \textbf{Min. Date} \\ %
\midrule
     Bitcoin &   1,349,572 & 6,461,916 &   7,811,488 & 2010-11-09 \\ %
    Ethereum &     336,850 & 1,076,066 &   1,412,916 & 2013-12-29\\ %
      Tether &      27,130 &   256,382 &     283,512 & 2012-06-13\\ %
         BNB &     169,854 &   291,180 &     461,034 & 2012-06-15\\ %
    USD Coin &       1,779 &     8,196 &       9,975 & 2012-08-10\\ %
         XRP &     122,419 &   563,620 &     686,039 & 2013-02-08\\ %
  Terra LUNA &       6,273 &     4,109 &      10,382 & 2020-08-19\\ %
     Cardano &      98,872 &   425,109 &     523,981 & 2016-12-09\\ %
      Solana &      52,277 &   175,008 &     227,285 & 2018-03-31\\ %
   Avalanche &      12,877 &    29,976 &      42,853 & 2012-02-11\\ %
 Binance USD &         522 &       938 &       1,460 & 2017-12-16\\ %
    Polkadot &         176 &        14 &         190 & 2019-02-19\\ %
    Dogecoin &     306,303 & 1,074,124 &   1,380,427 & 2013-12-03\\ %
    TerraUSD &         651 &     1,197 &       1,848 & 2020-11-20\\ %
   Shiba Inu &      56,711 &    44,925 &     101,636 & 2013-12-14\\ %
\bottomrule
\end{tabular}%
\caption{Overview of the top 15 cryptocurrencies in our dataset. (Max date is 2022-08-31 for all.)} %
\label{cryptocurrency_specific_stats_overview}
\end{table}

\section{Cryptocurrency-Level Analysis}

Next, we study the temporal evolution of posts mentioning specific cryptocurrencies.
More specifically, we focus on: a) studying the evolution of cryptocurrency-specific activity over time; 
b) performing an inter-community activity analysis; and 
c) investigating the relationship between cryptocurrency-specific activity and the price of a particular cryptocurrency using \textit{cross-correlation}.

To do so, we look for mentions of the cryptocurrency name in all submissions and comments in our dataset.
We do not search for cryptocurrency symbols as most of them match common words, thus leading to a large number of false positives.
If a submission or comment mentions more than one cryptocurrency, we include it in the sets of posts of all the cryptocurrencies it mentions.
Table~\ref{cryptocurrency_specific_stats_overview} reports the statistics for the top 15 cryptocurrencies (as per their market cap) in our dataset. %

\subsection{Cryptocurrency-specific Activity}
We analyze the evolution of submissions and comments mentioning each cryptocurrency over time.
Figure~\ref{fig:crypto_monthly_posts} reports the evolution of the monthly {\em normalized} number of submissions and comments mentioning a given coin along with its normalized price.
We also perform changepoint detection in the price of the cryptocurrency to investigate any possible correlations between the trends in the price and the posts mentioning it.
To ease presentation, we only discuss %
the most interesting findings. 

\begin{figure*}[t!]
\centering
\includegraphics[width=.9\linewidth]{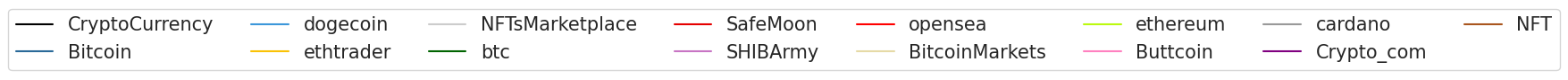}

\includegraphics[width=0.49\linewidth]{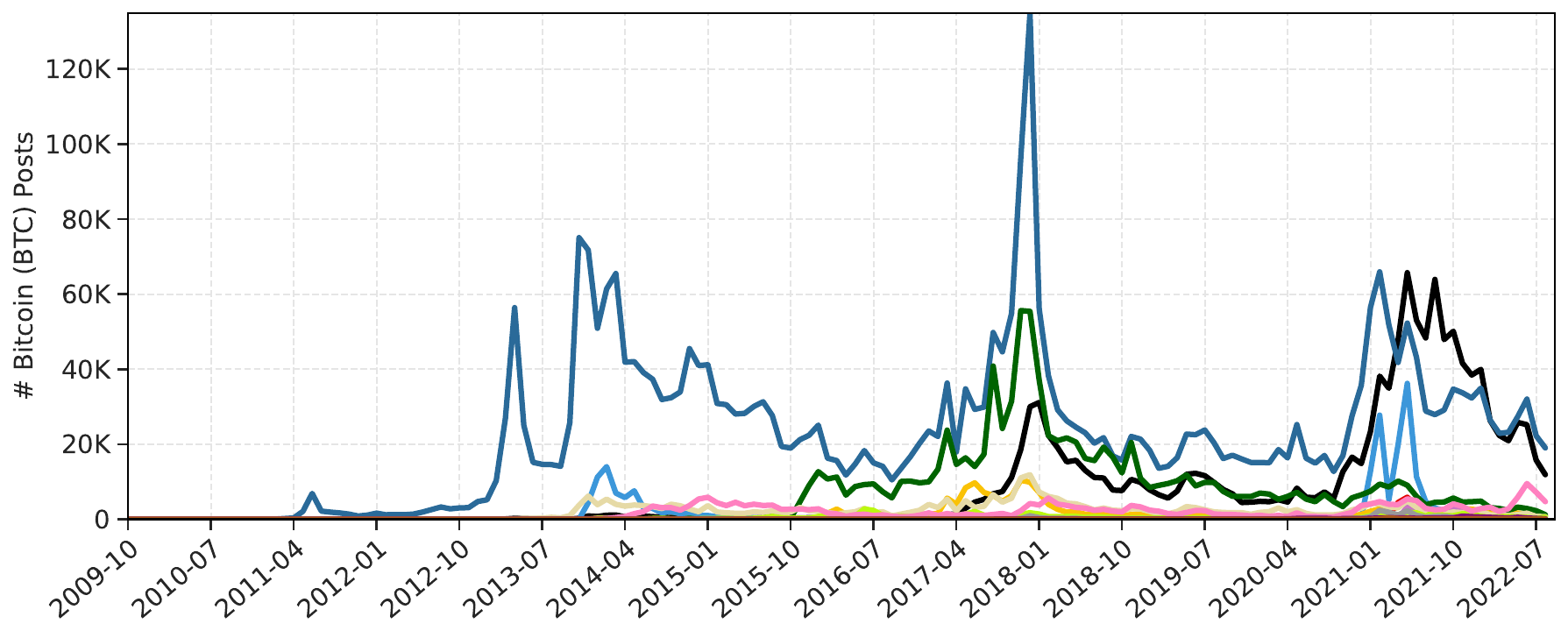}
\includegraphics[width=0.49\linewidth]{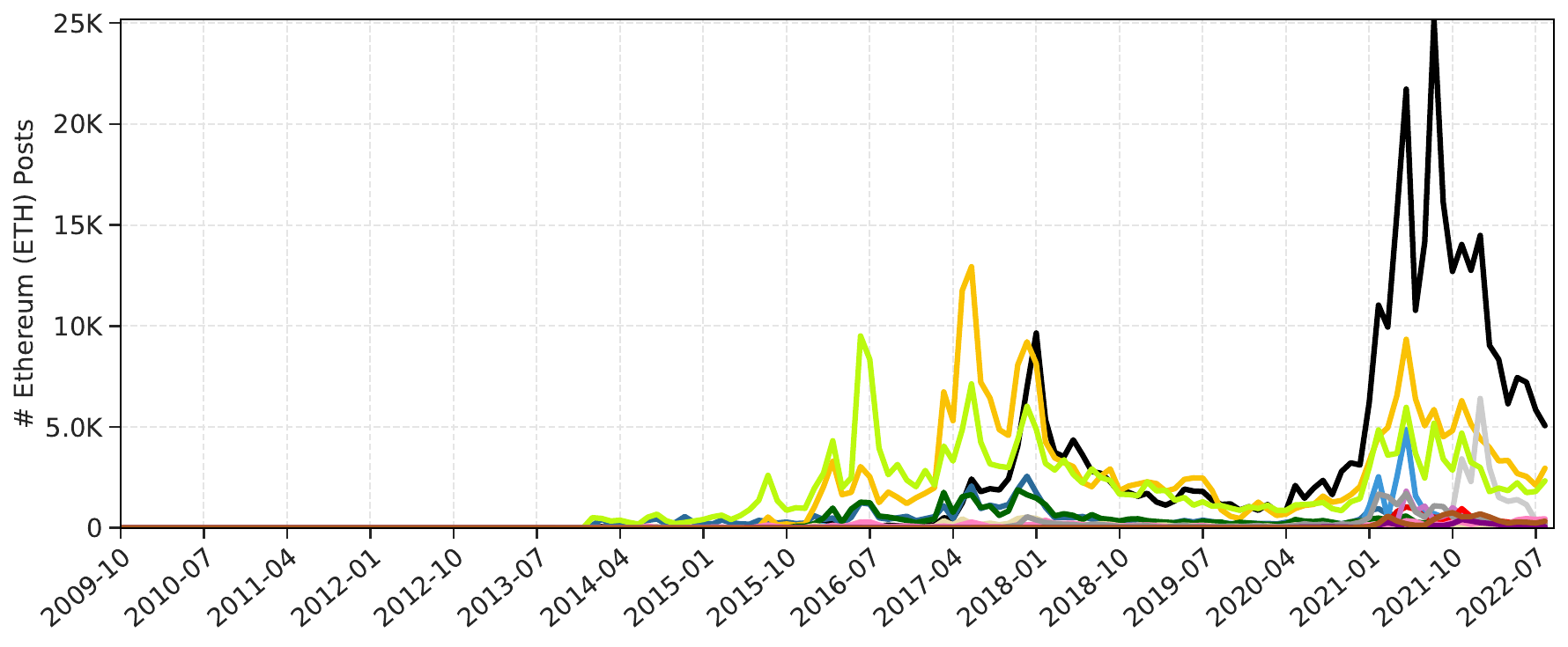}
\includegraphics[width=0.49\linewidth]{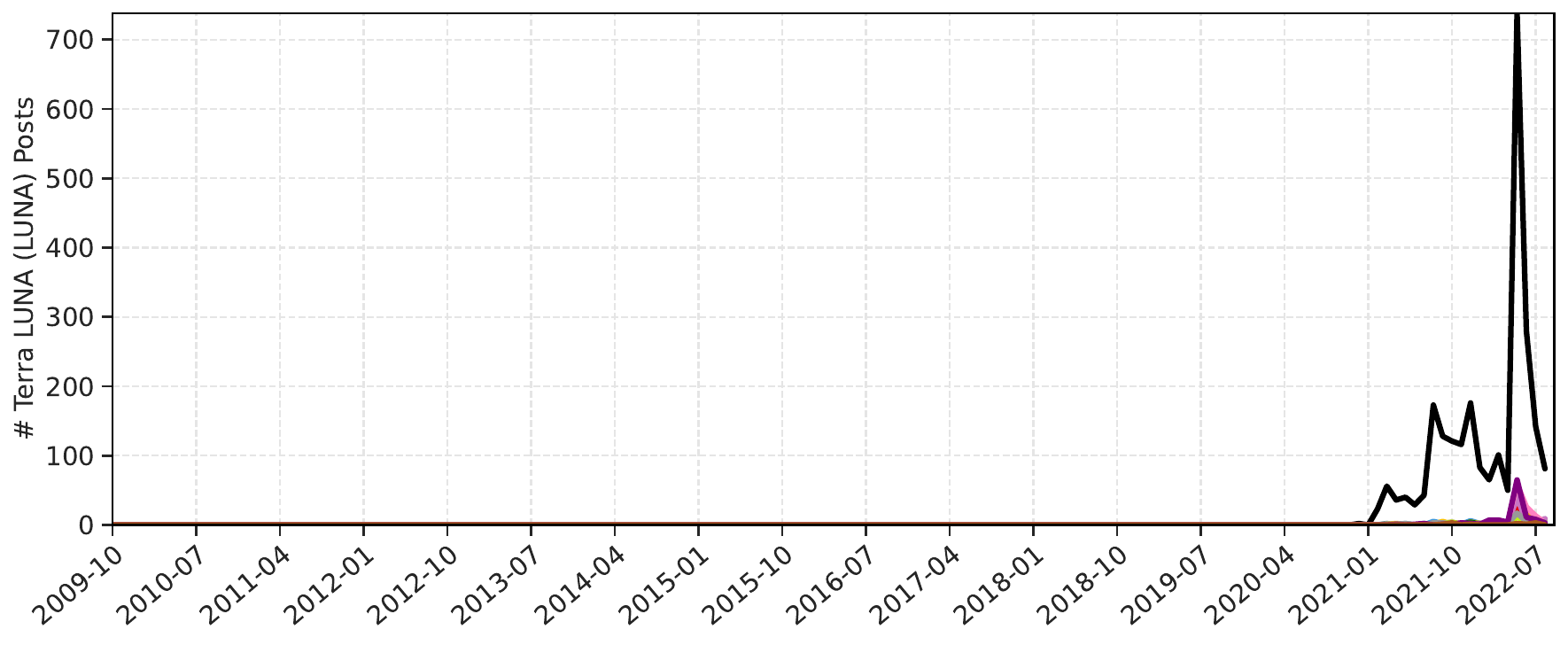}
\includegraphics[width=0.49\linewidth]{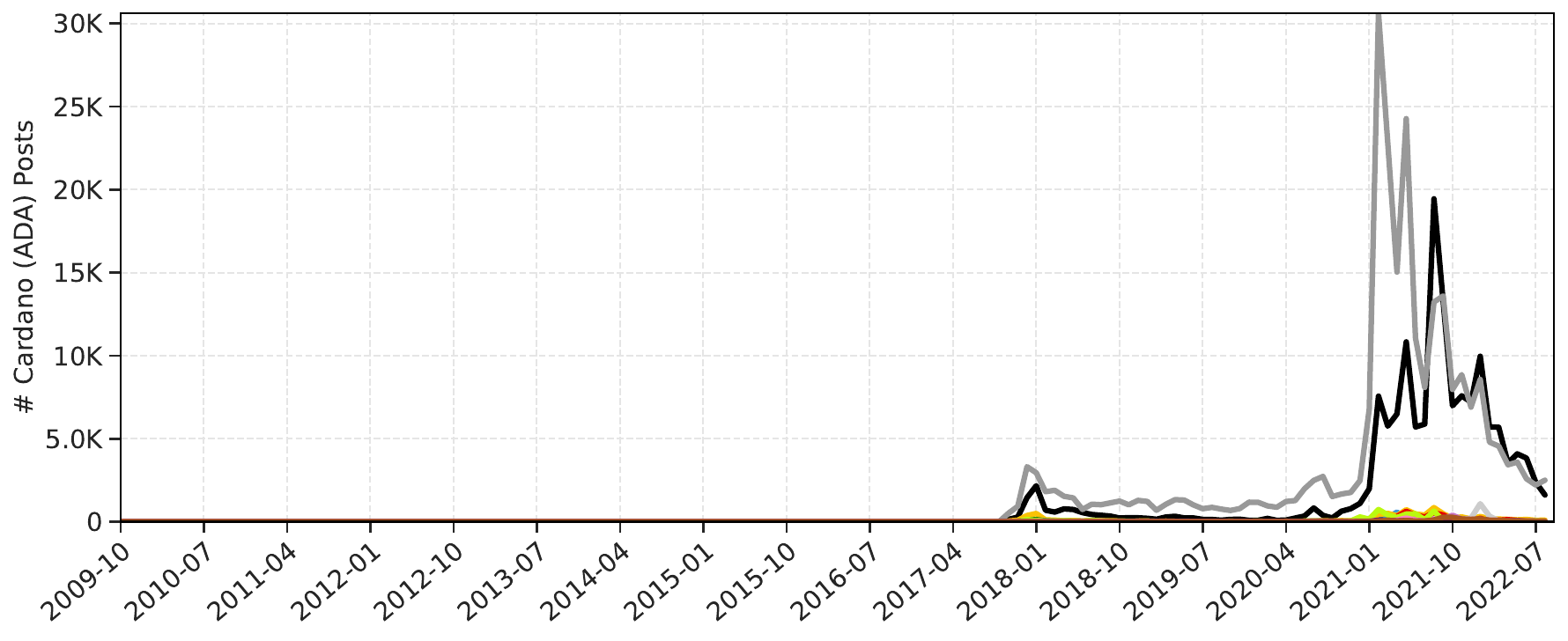}
\includegraphics[width=0.49\linewidth]{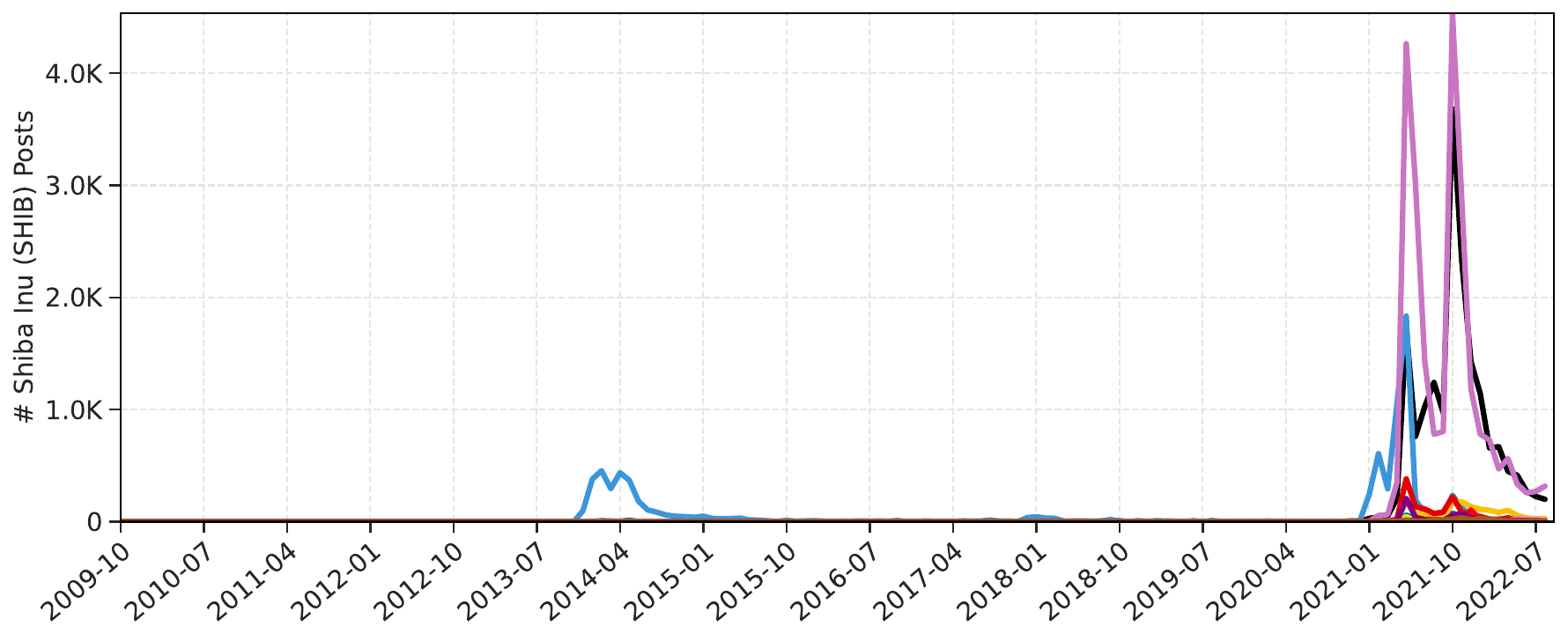}
\includegraphics[width=0.49\linewidth]{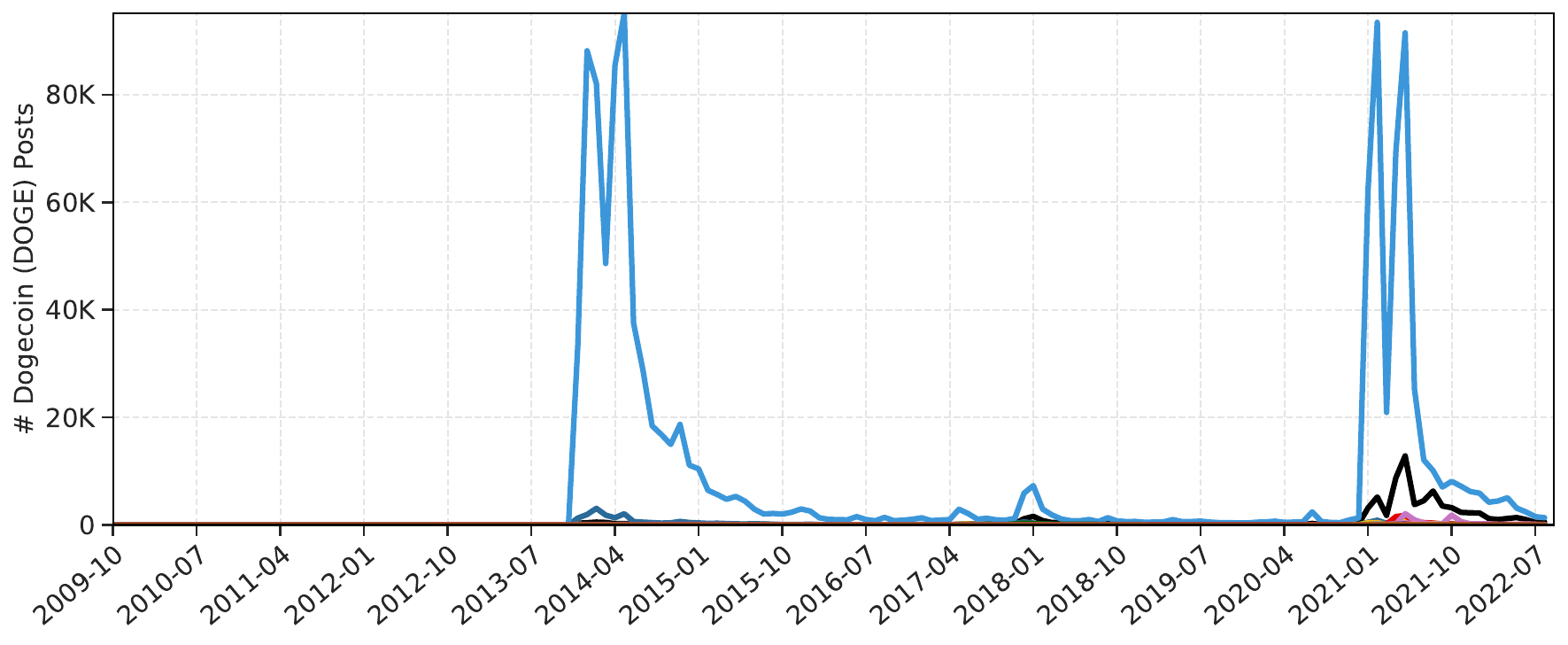}

\caption{Temporal evolution of the monthly absolute number of posts mentioning cryptocurrencies in the top 15 subreddits in our dataset.}
\label{fig:crypto_subreddits_monthly_posts}
\end{figure*}

\descr{Bitcoin (BTC).}
Bitcoin is the first decentralized digital currency to gain widespread popularity.
It was introduced in 2008 with Satoshi Nagatomo's whitepaper~\cite{nakamoto2008bitcoin}. 
In our dataset, we find the first ever post mentioning the term `bitcoin' on November 9, 2010.
Looking at Figure~\ref{fig:crypto_monthly_posts}, we find that Reddit users on cryptocurrency-related communities started discussing Bitcoin long before the large increases in the price of Bitcoin. 
However, we do find two changepoint events detected in the Bitcoin price aligned with the two sharp increases in Bitcoin-related activity in December 2018 and in the first half of 2021.
This suggests a relation between the activity increase and an increased interest in the cryptocurrency market in general, as spikes also happen when Bitcoin hit \$1T in market value in February 2021.

\descr{Ethereum (ETH).} Ethereum is the second most popular cryptocurrency based on the total number of posts mentioning the term `ethereum' in our dataset.
The first posts mentioning it appear in late 2013, roughly around the time Vitalik Buterin published the Ethereum's  whitepaper~\cite{buterin2014next}.
However, Ethereum-related activity starts rising only after the project's official launch on July 20, 2015.
From then on, the volume of submissions and comments mentioning Ethereum is largely aligned with its price.
Finally, Ethereum-related activity seems to take off in 2021, which coincides with the launch of the Ethereum Improvement Proposal (EIP) 1559~\cite{ethereumfork2021cnbc}.

\descr{Dogecoin (DOGE).}
Originally started as a joke, Dogecoin made its debut in December 2013, swiftly capturing the attention of the cryptocurrency community as confirmed by the high number of submissions and comments referencing Dogecoin.
A steady decline then follows this initial surge in activity lasting for several years.
Overall, peaks in activity can be primarily attributed to public endorsements of prominent figures such as Elon Musk; their endorsement generates substantial attention and participation within the Dogecoin community, consequently contributing to the heightened enthusiasm and activity surrounding it.

\descr{Terra LUNA (LUNA) and TerraUSD (UST).}
Our analysis highlights a remarkable rise and fall of Terra LUNA and TerraUSD discussions within a relatively short span of time.
The Terra network was introduced in January 2018 for e-commerce payment applications. 
In September 2020, TerraUSD was launched as an algorithmic stablecoin on the Terra blockchain.\footnote{An algorithmic stablecoin operates on the principle of using algorithms and smart contracts to (attempt to) maintain a stable value, distinguishing it from other types of cryptocurrencies that exhibit price volatility.} 
We then observe a surge in the value of Terra LUNA during the subsequent months of 2021.
Coinciding with this surge, we also see a large volume of posts referencing both Terra LUNA and TerraUSD.
However, a rapid escalation in the number of posts mentioning both Terra LUNA and TerraUSD occurs after May 8, 2022, which closely follows the collapse of the Terra Network starting the day before~\cite{lunacrush2022forbes}.

\descr{Decentraland (MANA).}
Decentraland is a blockchain-based endeavor that aims to establish a ``metaverse'' or virtual world, open and accessible to individuals at large.
The first submissions and comments referencing it date back to 2017, coinciding with the initial stages of development of the 
project.
This early interaction signifies the growing awareness and interest surrounding Decentraland within the community.
Next, following overall increased interest in the NFT ecosystem, Decentraland experiences a large surge in activity.
Notably, a substantial spike in engagement occurs in March 2021, followed by another in October 2021.
The former is accompanied by a 305\%  month-over-month increase in the price of Decentraland.

\subsection{Inter-Community Activity}
Next, in Figure~\ref{fig:crypto_subreddits_monthly_posts}, we report the number of monthly posts mentioning Bitcoin, Ethereum, Dogecoin, Terra LUNA, Cardano, Shiba Inu, and Dogecoin in the top 15 cryptocurrency-related subreddits over time.
This highlights which subreddits the top cryptocurrencies are being discussed on the most and lets us identify any subreddits devoted to discussing specific coins. %
Again, we only discuss the cryptocurrencies with the most interesting findings and omit the rest due to ease presentation.
Unsurprisingly, /r/Bitcoin, /r/btc, and /r/CryptoCurrency are the primary subreddits for Bitcoin-related discussions.
The highest number of posts mentioning Bitcoin is on /r/Bitcoin in the last quarter of 2017. 
This suggests a meaningful jump in Bitcoin-related activity, potentially driven by market dynamics and the heightened public interest in cryptocurrencies during that period.
Next, we find that Ethereum is mostly discussed in /r/ethereum, /r/ethtrader, and /r/CryptoCurrency, with a substantial amount of Ethereum-related activity within the /r/CryptoCurrency subreddit throughout 2021. 
This surge in Ethereum-related activity aligns with the release of Ethereum 2.0, a highly anticipated upgrade to the Ethereum network. %

\begin{figure*}[t!]
\centering
\begin{subfigure}[b]{0.245\textwidth}
	\includegraphics[width=\linewidth]{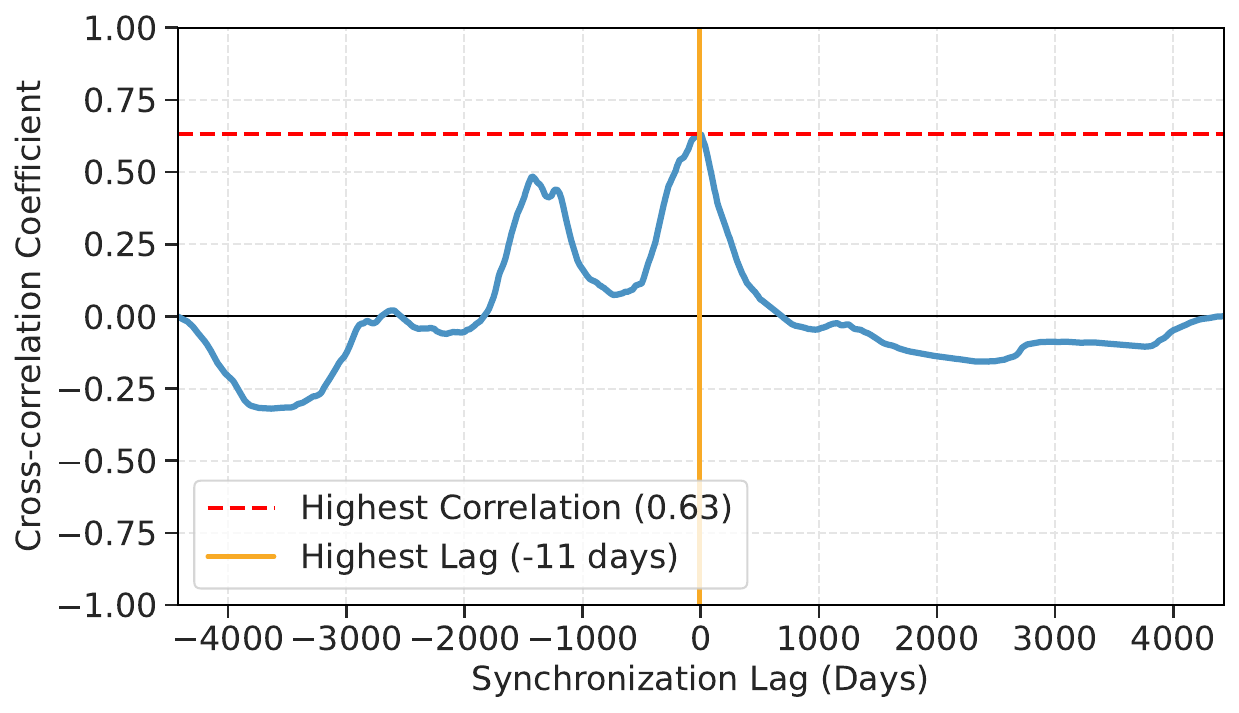}
    \caption{Bitcoin (BTC)}\myspace\myspace
\end{subfigure}
\begin{subfigure}[b]{0.245\textwidth}
	\includegraphics[width=\linewidth]{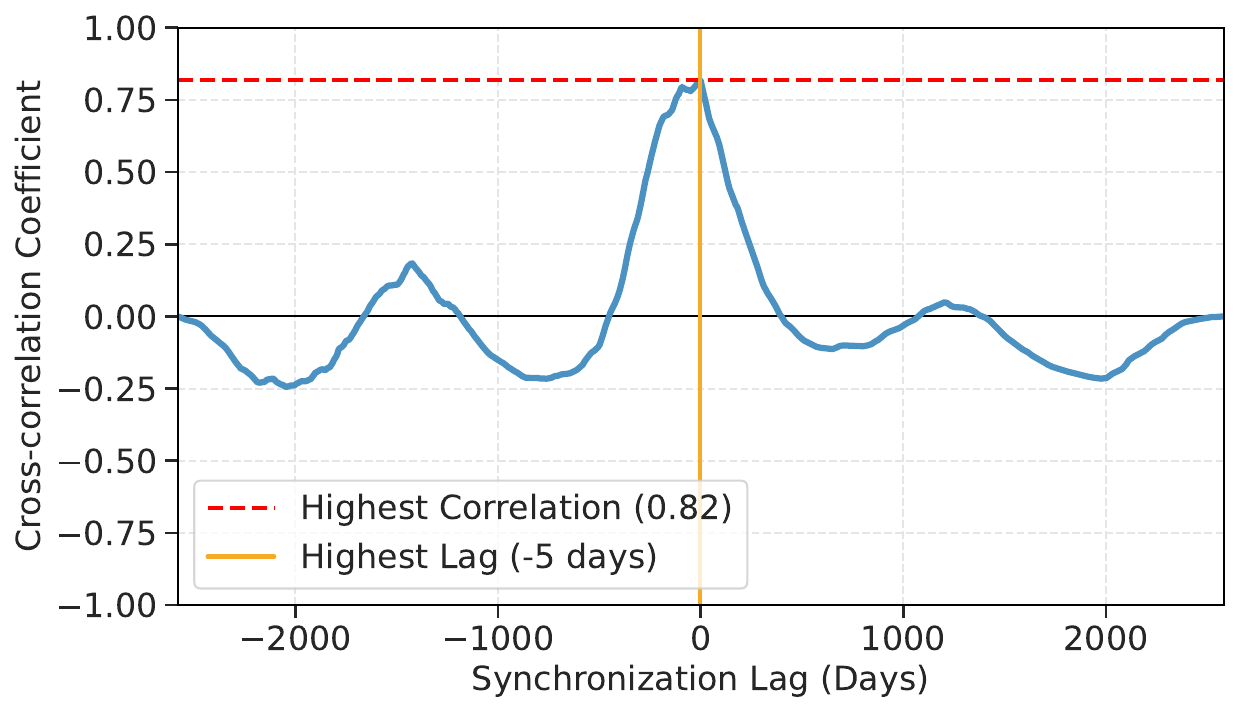}
    \caption{Ethereum (ETH)}\myspace\myspace
\end{subfigure}
\begin{subfigure}[b]{0.245\textwidth}
	\includegraphics[width=\linewidth]{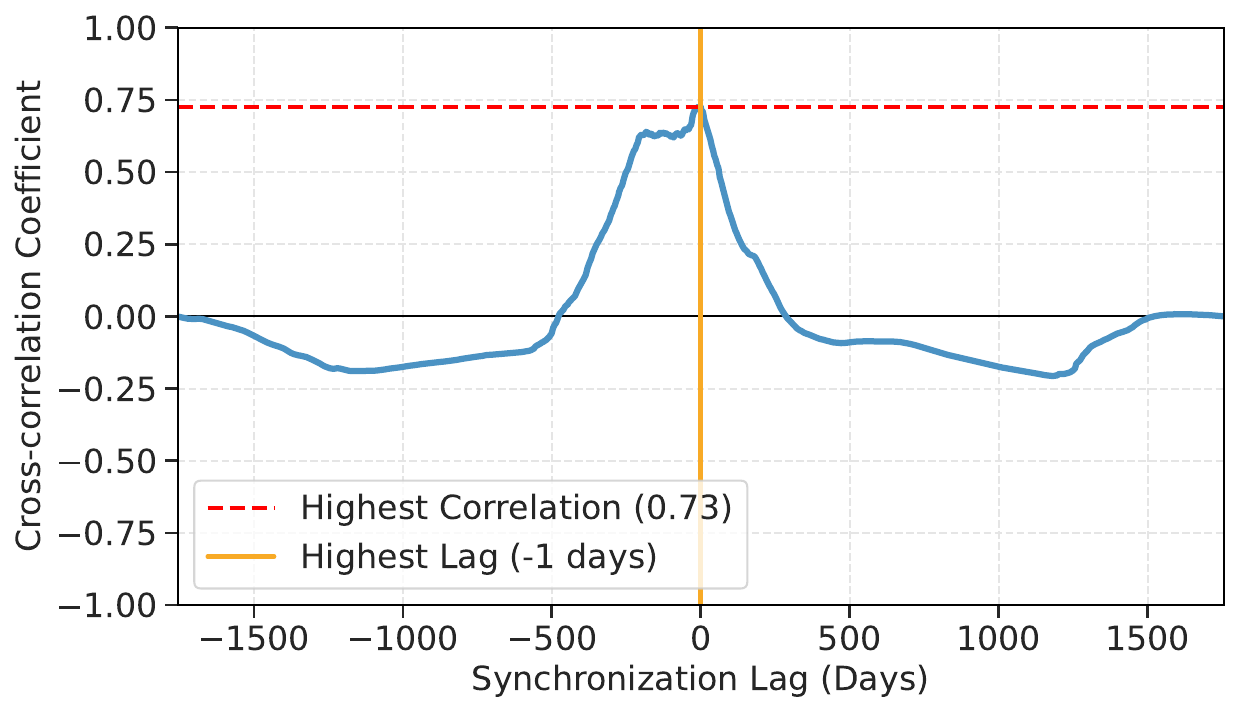}
    \caption{BNB (BNB)}\myspace\myspace
\end{subfigure}
\begin{subfigure}[b]{0.245\textwidth}
	\includegraphics[width=\linewidth]{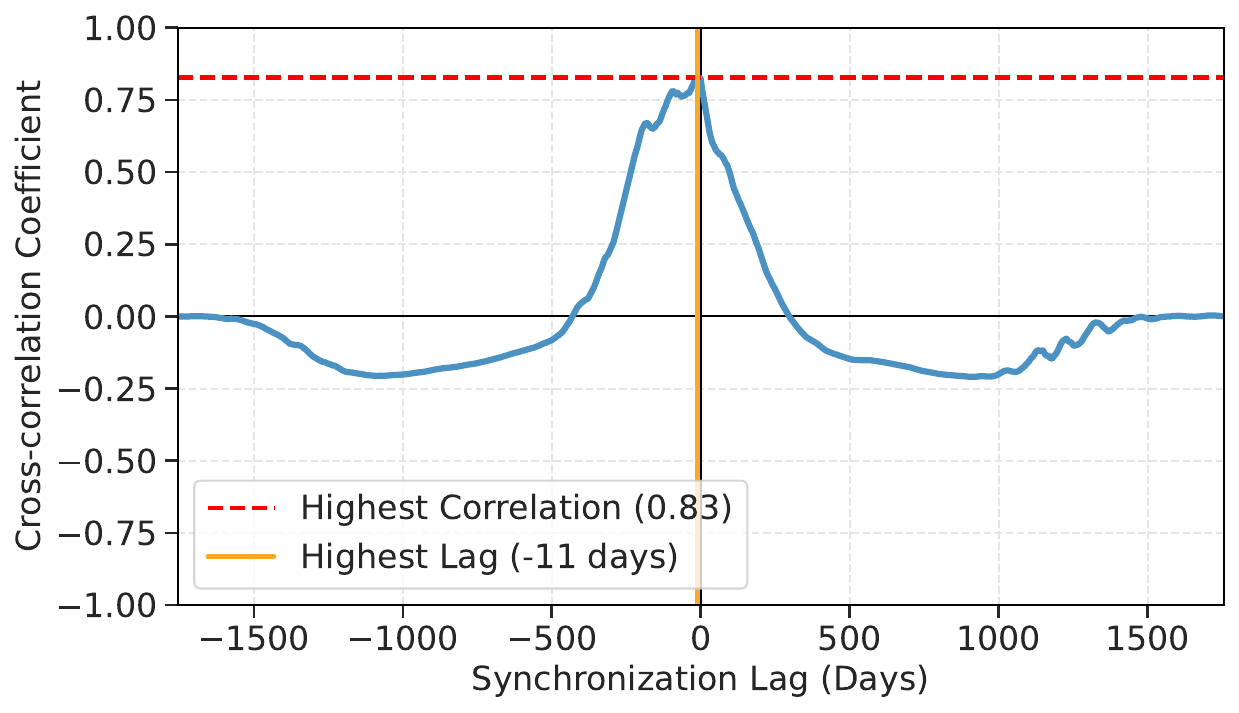}
    \caption{Cardano (ADA)}\myspace\myspace
\end{subfigure}
\begin{subfigure}[b]{0.245\textwidth}
	\includegraphics[width=\linewidth]{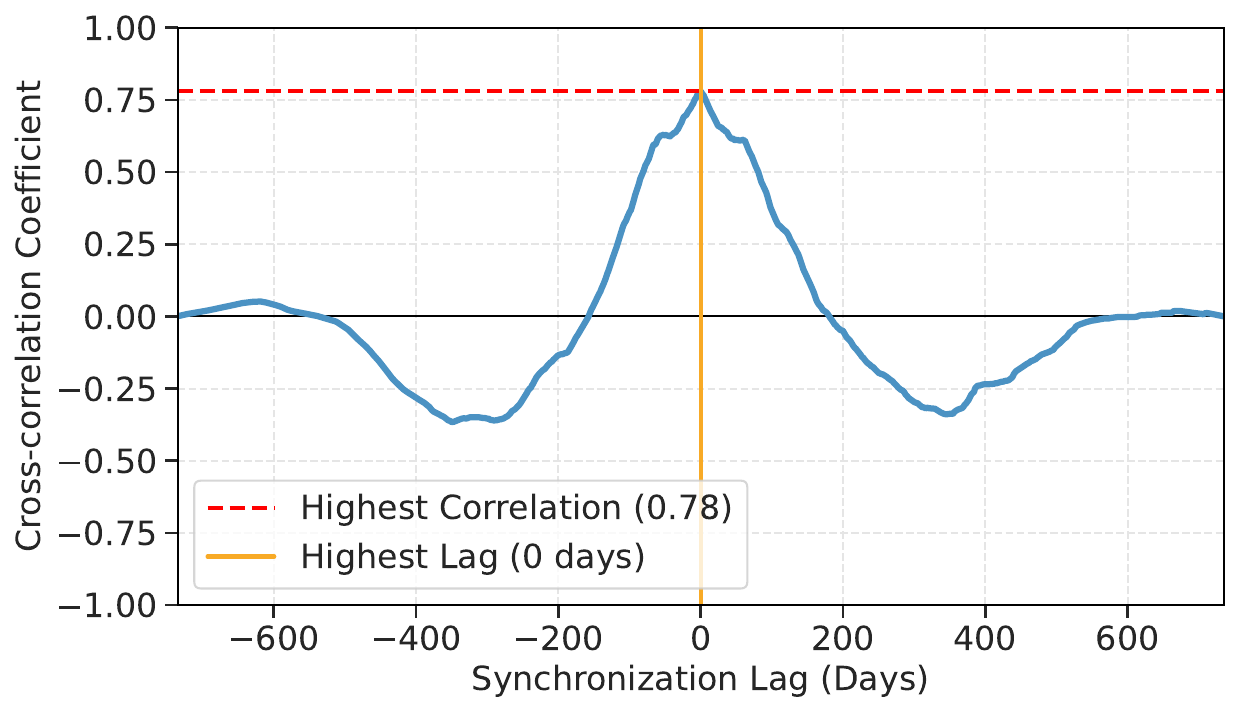}
    \caption{Solana (SOL)}\myspace\myspace
\end{subfigure}
\begin{subfigure}[b]{0.245\textwidth}
	\includegraphics[width=\linewidth]{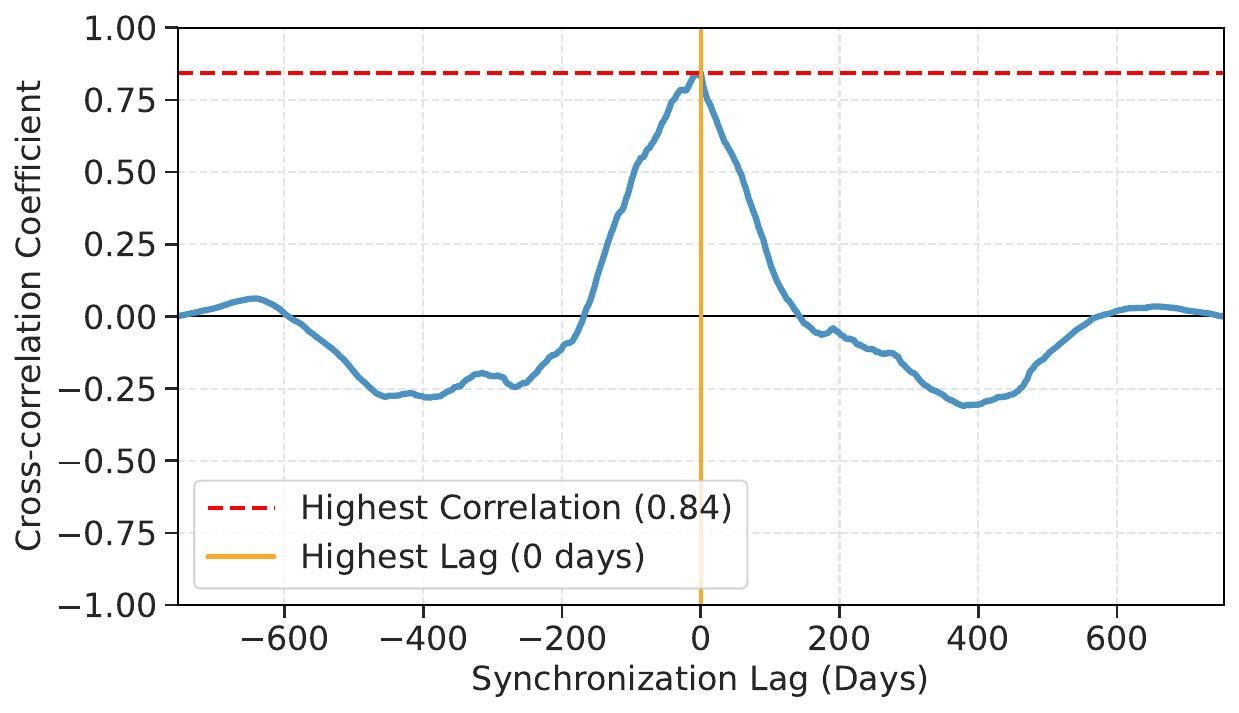}
    \caption{Avalanche (AVAX)}\myspace\myspace
\end{subfigure}
\begin{subfigure}[b]{0.245\textwidth}
	\includegraphics[width=\linewidth]{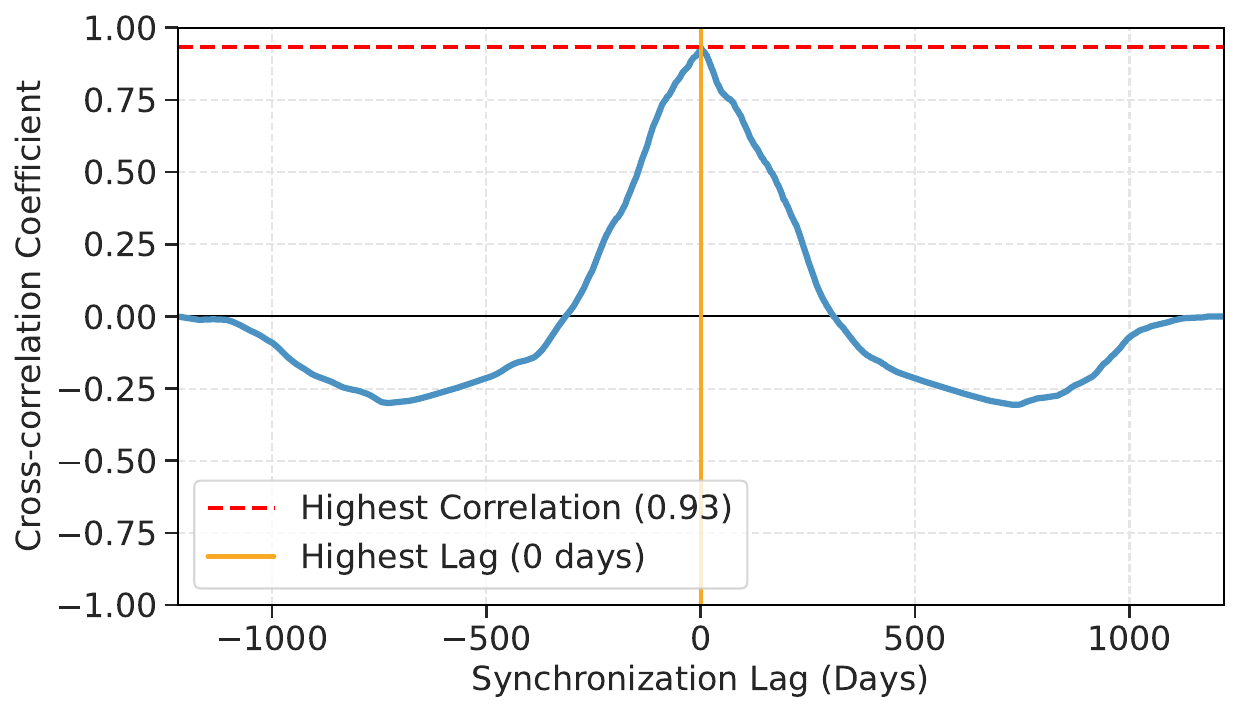}
    \caption{Polygon (MATIC)}\myspace\myspace
\end{subfigure}
\begin{subfigure}[b]{0.245\textwidth}
	\includegraphics[width=\linewidth]{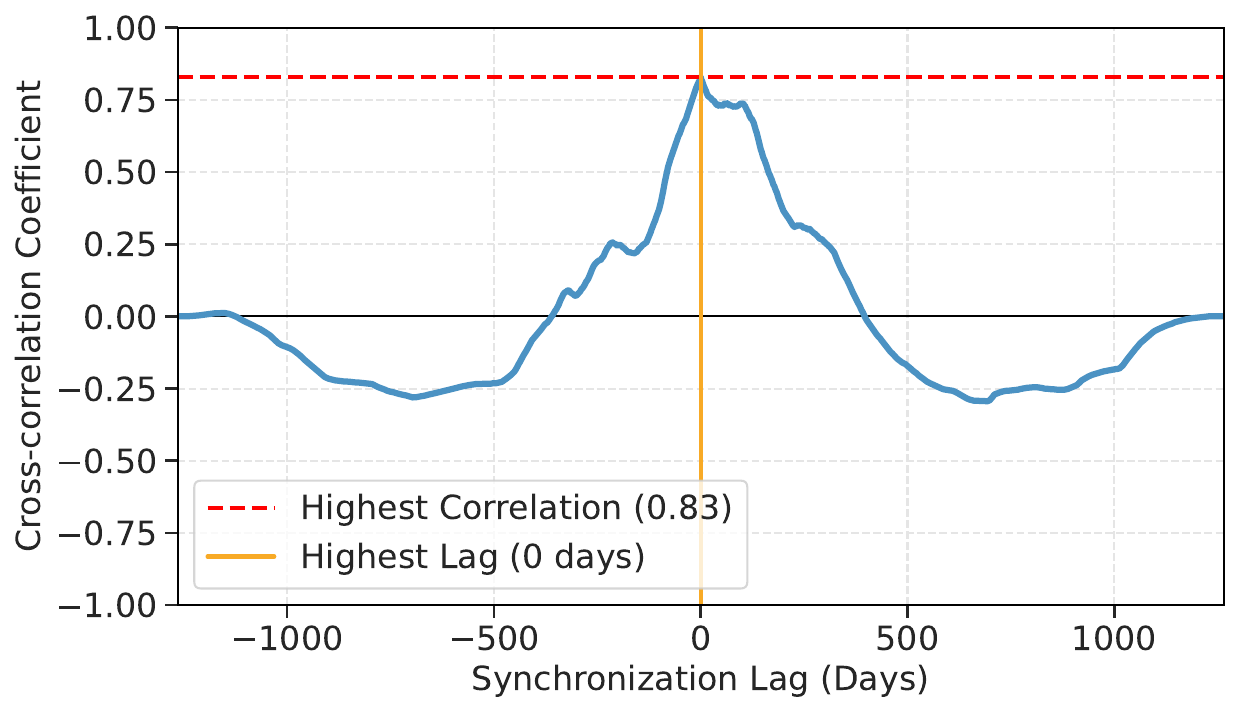}
    \caption{Cosmos (ATOM)}\myspace\myspace
\end{subfigure}
\begin{subfigure}[b]{0.245\textwidth}
	\includegraphics[width=\linewidth]{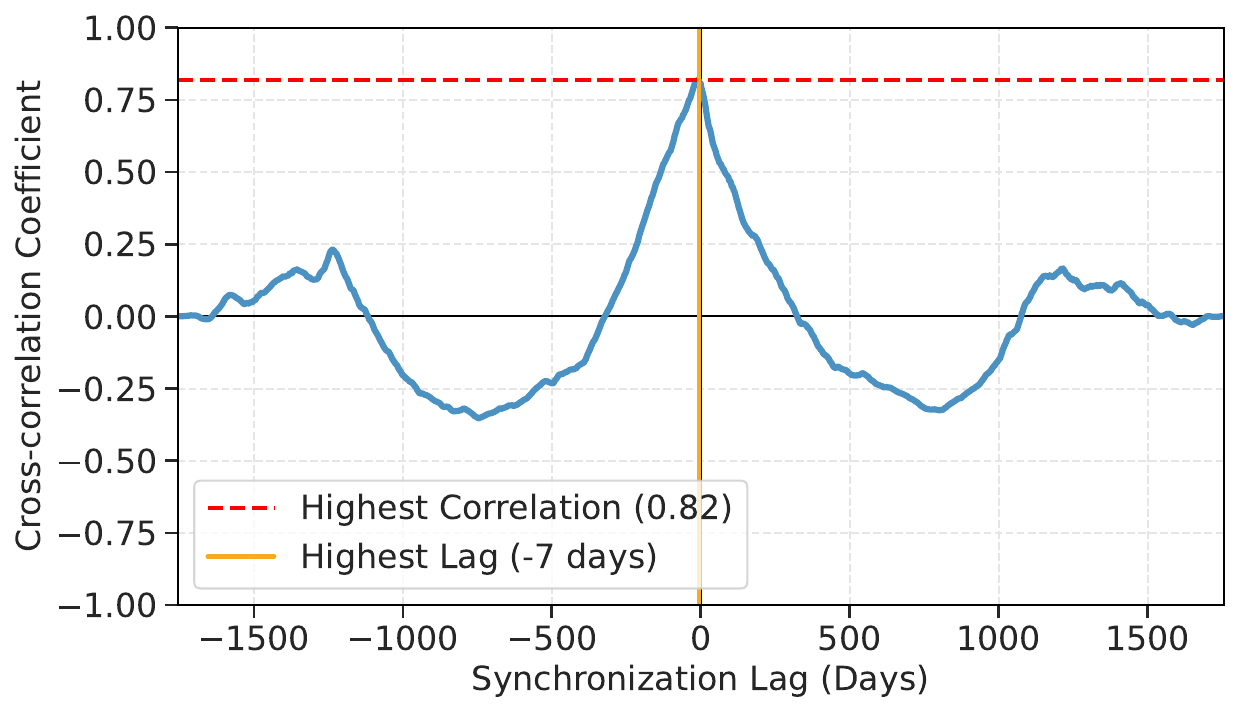}
    \caption{Monero (XMR)}\myspace\myspace
\end{subfigure}
\begin{subfigure}[b]{0.245\textwidth}
	\includegraphics[width=\linewidth]{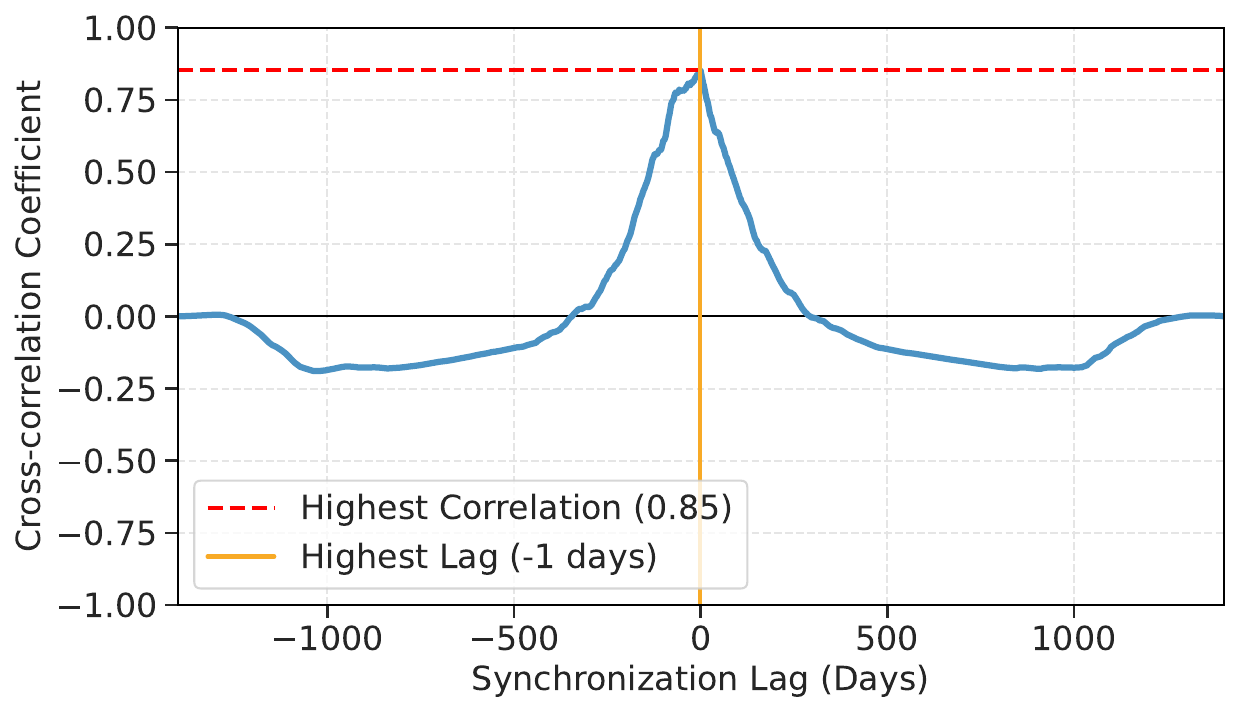}
    \caption{Fantom (FTM)}\myspace\myspace
\end{subfigure}
\begin{subfigure}[b]{0.245\textwidth}
	\includegraphics[width=\linewidth]{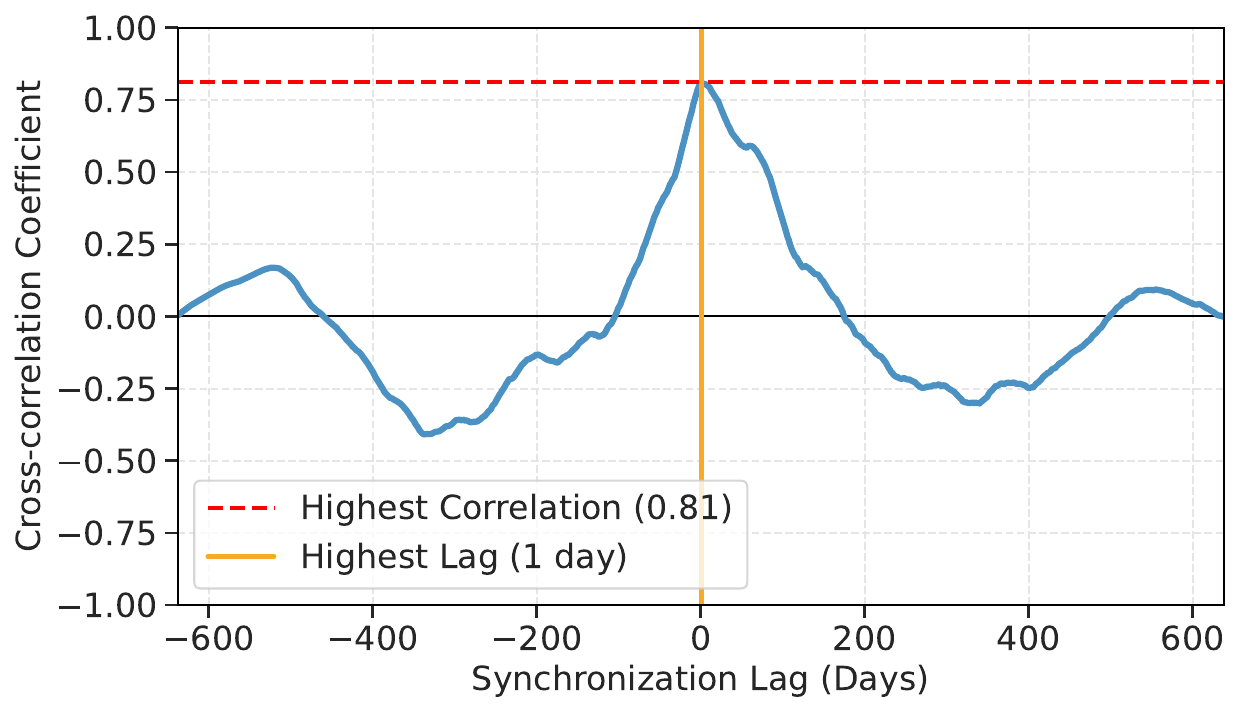}
    \caption{Helium (HNT)}\myspace\myspace
\end{subfigure}
\begin{subfigure}[b]{0.245\textwidth}
	\includegraphics[width=\linewidth]{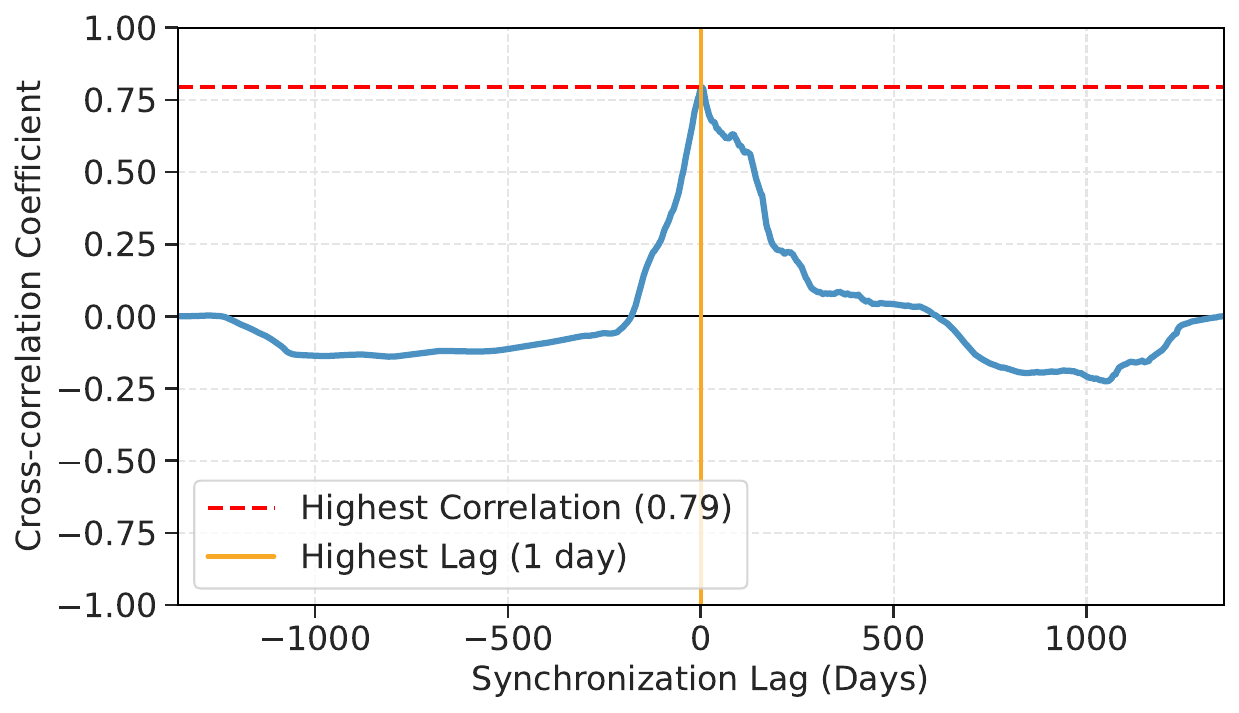}
    \caption{Cronos (CRO)}\myspace\myspace
\end{subfigure}	
\caption{Cross-correlation analysis of the daily number of posts mentioning a cryptocurrency and its price.}
\label{fig:crypto_posts_price_crosscorrelation}
\end{figure*}

The /r/CryptoCurrency subreddit is also the predominant platform for Terra LUNA, %
with a large surge occurring around May 2022, coinciding with the collapse of the LUNA network~\cite{lunacrush2022forbes}.
A relatively smaller portion of discussions can also be found on /r/Crypto\_com.

Expectedly, /r/cardano is the primary hub for discussions centered around Cardano, with a noteworthy surge in activity after 2020, with February 2021 having the highest number of posts. 
A substantial proportion of Cardano discussions is also found within /r/CryptoCurrency, which seems to experience a comparatively delayed increase in Cardano-related activity, even more so in August 2021.
Also, Shiba Inu garners substantial attention on /r/SHIBArmy, with some activity also occurring on /r/CryptoCurrency, with a synchronized increase in engagement during 2021 on both subreddits.
Similarly, the bulk of discussions related to Dogecoin is concentrated on /r/dogecoin, with two interesting peaks.
The first dates back to 2013--2014, coinciding with the initial launch of the coin, suggesting a surge of interest during its early stages.
The second one materializes in 2021, when Dogecoin experiences a notable upswing in activity.
As mentioned before, this can be attributed to the involvement of prominent figures like Elon Musk.

\subsection{Price Synchronization}
Finally, we use signal processing techniques to study possible correlations and synchronization between the daily close price of a cryptocurrency and the daily number of posts mentioning it %
using \textit{cross-correlation}.

Cross-correlation slides one signal with respect to the other and calculates the dot product (i.e., the \textit{matching}) between the two signals for each possible lag.
The estimated lag is the one maximizing the matching between the signals.
This allows us to uncover relationships between discussions around a cryptocurrency and changes in its price. 
Such correlations could occur if changes in price are followed by increased activity around that cryptocurrency or vice versa.
Naturally, correlation does not imply causation; rather, our cross-correlation analysis provides insights into the temporal relationship between the variables we are examining.

In Figure~\ref{fig:crypto_posts_price_crosscorrelation}, we plot the cross-correlation between each cryptocurrency price and the daily number of posts mentioning that particular cryptocurrency for all different synchronization lags.
A lag of 1 means a synchronization lag of 1 day.
Once again, we limit our discuss to the most interesting findings. %
Note that, for each cryptocurrency, we consider only submissions and comments created {\em after} the date for which we have historical price data.

\descr{Bitcoin (BTC).} For Bitcoin, we find a cross-correlation of 0.62, which is considered a ``moderate'' positive correlation, between the Bitcoin daily price and the daily number of posts mentioning Bitcoin.
The fact that the highest synchronization lag is -11 days indicates that the number of posts tends to lead changes in the Bitcoin price by 11 days.

\descr{Ethereum (ETH).}
Interestingly, for Ethereum we find a 0.82 correlation, which indicates a ``strong'' positive correlation between price and posts mentioning Ethereum.
The highest lag of -5 days suggests that changes in the number of posts tend to anticipate price changes by 5 days.

\descr{BNB (BNB).}
We also find a relatively strong correlation of 0.73 for BNB, with a highest lag of -1 day.

\descr{Cardano (ADA).}
Similar to Bitcoin, we find a highest lag of -11 days for Cardano.
However, a 0.83 correlation coefficient here suggests a very ``strong'' positive correlation. %

\descr{Additional Coins.} We also find relatively ``strong'' positive correlations between the changes in the number of posts and the price of Solana (0.78), Avalanche (0.84), Polygon (0.93), and Cosmos (0.82), with highest lag for all of them of 0 days.
This means that there is an immediate relationship between discussions and price, and that changes in both variables occur within the same day.
A ``strong'' positive correlation between the daily number of posts and the price movement is also observed for Monero (0.82 with a -7 days lag) and Fantom (0.85 with a -1 days lag).
For Cronos and Helium, we find a correlation coefficient of 0.79 and 0.81, respectively, and a positive highest lag of 1 day.

Overall, while there is a strong relationship between the number of posts and the price of certain cryptocurrencies, it does not necessarily mean that the number of posts directly causes price changes or vice versa. 
Other factors, such as market dynamics, news events, or investor behavior, could also play an important role. 
As a result, we call for future work analyzing additional factors and gaining a more comprehensive understanding of the relationship between cryptocurrency prices and their social network presence.

\begin{table}[t]
\centering
\small
\setlength{\tabcolsep}{4pt}
\begin{tabular}{lrrrr}
\toprule
\textbf{Currency}  & \textbf{Lag}   &  \textbf{CC Score} &  \textbf{Portfolio Value} &  \textbf{Return} \\
\midrule
Solana            & -5  & 0.78 & \$\numprint{31916439} & 3091.64\% \\
Polygon           & 0   & 0.93 & \$\numprint{17983091} & 1698.31\% \\
Axie Infinity     & -35 & 0.57 & \$\numprint{11952354} & 1095.24\% \\
TRON              & -1  & 0.72 & \$\numprint{8184047}  & 718.40\%  \\
Fantom            & -1  & 0.85 & \$\numprint{5855175}  & 485.52\%  \\
Hedera            & 0   & 0.79 & \$\numprint{5811347}  & 481.13\%  \\
Decentraland      & 0   & 0.72 & \$\numprint{4286751}  & 328.68\%  \\
Ethereum          & -5  & 0.82 & \$\numprint{3250647}  & 225.06\%  \\
THORChain         & 0   & 0.58 & \$\numprint{2745402}  & 174.54\%  \\
Avalanche         & 0   & 0.84 & \$\numprint{2416548}  & 141.65\%  \\
Cardano           & -11 & 0.83 & \$\numprint{2118245}  & 111.82\%  \\
Vechain           & -2  & 0.51 & \$\numprint{1854393}  & 85.44\%   \\
The Sandbox       & -1  & 0.76 & \$\numprint{1357773}  & 35.78\%   \\
Zcash             & -1  & 0.56 & \$\numprint{1326302}  & 32.63\%   \\
Litecoin          & -6  & 0.54 & \$\numprint{1275137}  & 27.51\%   \\
BNB               & -1  & 0.73 & \$\numprint{1235578}  & 23.56\%   \\
Shiba Inu         & 0   & 0.54 & \$\numprint{998057}   & -0.19\%   \\
Tezos             & -20 & 0.59 & \$\numprint{892997}   & -10.70\%  \\
Ethereum Classic  & -2  & 0.53 & \$\numprint{783639}   & -21.64\%  \\
Cosmos            & 0   & 0.83 & \$\numprint{714024}   & -28.60\%  \\
Bitcoin           & -11 & 0.63 & \$\numprint{550214}   & -44.98\%  \\
XRP               & -3  & 0.57 & \$\numprint{488271}   & -51.17\%  \\
Chainlink         & 0   & 0.75 & \$\numprint{448449}   & -55.16\%  \\
Algorand          & 0   & 0.74 & \$\numprint{440951}   & -55.90\%  \\
Bitcoin Cash      & -1  & 0.76 & \$\numprint{402104}   & -59.79\%  \\
Stellar           & -3  & 0.72 & \$\numprint{383168}   & -61.68\%  \\
Monero            & -7  & 0.82 & \$\numprint{370401}   & -62.96\%  \\
Wrapped Bitcoin   & -72 & 0.54 & \$\numprint{266505}   & -73.35\%  \\
Filecoin          & -15 & 0.57 & \$\numprint{105790}   & -89.42\%  \\
Internet Computer & 0   & 0.61 & \$\numprint{100895}   & -89.91\% \\ \midrule
    \textbf{Total} &      \textbf{--} &  \textbf{--} &  \textbf{\$\numprint{110514693}} &     \textbf{268.38\%} \\      
\bottomrule
\end{tabular}
\caption{Backtesting results for the 30 cryptocurrencies with a $\leq 0$ lag and a cross-correlation score (CC) over 0.5, sorted by decreasing returns. Starting portfolio is \$1M for each currency.} 
\label{tab:backtesting_results_overview}
\end{table}

\revision{\subsection{Insights Evaluation}
To shed more light on our price synchronization analysis, we perform {\em backtesting}\footnote{Backtesting is a systematic empirical evaluation process analyzing historical market data and trading strategies to assess how well the latter would have performed in the past, offering insights into their potential profitability and risk management.} 
using the historical market data of the 30 cryptocurrencies for which we find a ``strong'' positive correlation between the changes in the number of posts and their market price.
}

\revision{
To do so, we use the following straightforward strategy.
For each cryptocurrency, we start our backtesting on January 1, 2021, and perform daily trading using the daily close price of the cryptocurrency until August 21, 2022.
To decide whether to buy, sell, or hold a certain position, we analyze the trend in the number of posts $k$ days ago, where $k$ is the largest lag from our cross-correlation analysis for each cryptocurrency. 
The trend in the number of posts each day is determined by comparing the current day's post count to the previous day's count.
Specifically, we \emph{buy} if the number of posts on a given day exceeds the previous day and \emph{sell} if it decreases.
If it stays the same, we \emph{hold} our position.}

\revision{
We trade each cryptocurrency independently, starting with an initial balance of \$1M for each, and we assume market orders exclusively.
Our daily strategy involves liquidating all available holdings when we decide to sell. 
Conversely, when we buy, we invest the entirety of our available cash account balance into the particular cryptocurrency.
We also factor in transaction fees at 0.1\%, similar to Binance (a prominent cryptocurrency exchange platform). %
}

\revision{Table~\ref{tab:backtesting_results_overview} reports the detailed results of our backtesting evaluation.
Backtesting on Solana yields the highest return (3,091.64\%), followed by Polygon (1,698.31\%) and TRON (718.40\%)
While we get roughly the same number of positive (16) and negative (14) returns vis-\`a-vis each coin, our strawman investment strategy yields an impressive +268.38\% total return. %
This denotes that even a simple trading strategy based on the trend of the daily number of posts can yield a considerable return.
While other factors play an important role in cryptocurrency changes, our backtesting results confirm a strong relationship between the number of posts and the price movement of popular cryptocurrencies.
}

\section{Emotion Analysis} %
In this section, we build and run an emotion detection classifier to identify primary emotions from users' posts. 
We do so to examine the temporal dynamics of users' emotions and how they relate to the cryptocurrency market.
We also explore the emotions specific to each community. %

\begin{figure*}[t!]
  \centering
    \includegraphics[width=1.25\columnwidth]{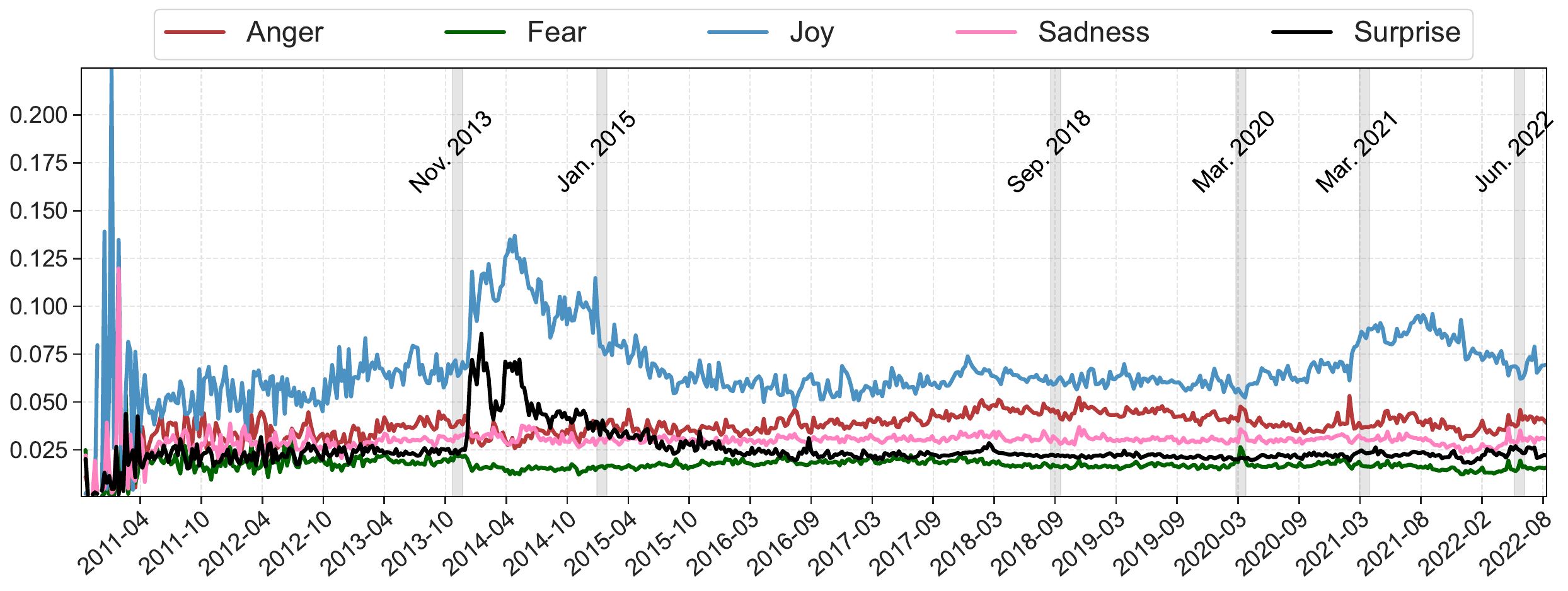}
  \caption{Temporal evolution of emotion scores. Shaded months represent when major cryptocurrency-related events occurred.}
\label{fig:emo_timeline}
\end{figure*}

\subsection{Emotion Detection}
\label{sec:emotion_detection}

\descr{Model.} To build an emotion detection model,
we rely on %
RoBERTa~\cite{liu2019roberta}, a pre-trained BERT-based model. %
We choose it following related work; e.g.,~\citet{murarka2021classification} build a multi-class mental illness Reddit classifier, showing it outperforms LSTM and BERT. 
Our work uses the RoBERTa architecture with 110M parameters and a hidden dimension size of 768.

\revision{We select five basic emotions from Ekman's emotion theory~\cite{ekman1992argument}: joy, sadness, anger, fear, and surprise. 
We believe these are relevant to decision-making, risk perception, and market sentiment, all arguably important in the context of the cryptocurrency market~\cite{breaban2017emotional,nguyen2014risk}.%
We also add the `neutral' emotion to represent neutral sentiments on social media discussions, following~\cite{araque2023emit}.}

\revision{RoBERTa was trained on a diverse dataset of 160 GB of data and demonstrated the ability to perform downstream tasks on new datasets using transfer learning~\cite{rajapaksha2021bert}. %
We develop the emotion detection classifier with the assumption that each post in our dataset corresponds to only one emotion and not multiple emotions.
We rely on an existing dataset with posts expressing one of six emotions to perform multi-class classification, collecting \numprint{407491} labeled samples from~\cite{saravia2018carer} and~\cite{demszky2020goemotions} and creating an 80/20 split for training and testing sets. 
We then fine-tune the model for two epochs using the same parameters as~\cite{murarka2021classification}, tuning the weights of all layers of the RoBERTa architecture instead of freezing any layers.
This yields 95.62\% accuracy on the test dataset.
We train our model using an NVIDIA GPU A100 with 80 GB of memory. 
}

\begin{table}[t!]
\centering
\small
\setlength{\tabcolsep}{1pt}
\resizebox{\columnwidth}{!}{%
\begin{tabular}{lrr@{}rrr@{}r}
\toprule      
\textbf{Subreddit} &  \textbf{\#Posts} &  \textbf{Joy} &  \textbf{~Sadness} & \textbf{Anger} & \textbf{Fear} & \textbf{~Surprise} \\
\midrule
 
 /r/CryptoCurrency & \numprint{27599717} & 0.08 & 0.03 & 0.05 & 0.02 & 0.02 \\
 
        /r/Bitcoin & \numprint{14168697} & 0.06 & 0.03 & 0.05 & 0.02 & 0.02 \\
        
	{/r/dogecoin} & \numprint{11686460} & 0.10 & 0.03 & 0.03 & 0.01 & $\uparrow$\textbf{0.04}  \\
       
      /r/ethtrader & \numprint{4831815} & 0.07 & 0.03 & 0.04 & 0.02 & 0.02 \\
      
{/r/NFTsMarketplace} & \numprint{4102847} & 0.10 & $\downarrow$\textbf{0.01} & $\downarrow$\textbf{0.01} & $\downarrow$\textbf{0.01} & 0.03  \\

            /r/btc & \numprint{3857648} & 0.05 & 0.03 & 0.05 & 0.02 & 0.02  \\
            
       /r/SafeMoon & \numprint{3119910} & 0.09 & 0.03 & 0.05 & 0.02 & 0.02  \\
       
       {/r/SHIBArmy} & \numprint{2121242} & 0.08 & 0.03 & 0.06 & 0.02 & $\downarrow$\textbf{0.02} \\
       
        /r/opensea & \numprint{2168988} & 0.10 & 0.01 & 0.01 & 0.01 & 0.03 \\
        
	{/r/BitcoinMarkets} & \numprint{2184587} & 0.06 & 0.03 & 0.04 & $\uparrow$\textbf{0.02} & 0.02 \\
 
       /r/ethereum & \numprint{1496575} & 0.06 & 0.03 & 0.03  & 0.02 & 0.02 \\
       
       {/r/Buttcoin} & \numprint{1625852} & $\downarrow$\textbf{0.05} & $\uparrow$\textbf{0.03} & $\uparrow$\textbf{0.07} & 0.02 & 0.03 \\
       
        /r/cardano & \numprint{1045334} & 0.09 & 0.03 & 0.03 & 0.02 & 0.02 \\
        
    /r/Crypto\_com & \numprint{1134590} & 0.08 & 0.04 & 0.04 & 0.02 & 0.02 \\
    
            {/r/NFT} & \numprint{730010} & $\uparrow$\textbf{0.13} & 0.02 & 0.02 & 0.00 & 0.03 \\
            
\bottomrule
\end{tabular}
}
\caption{Average emotion scores for the top 15 subreddits in our dataset $\uparrow$ and $\downarrow$ denote, respectively, highest and lowest scores.} %
\label{tab:subreddit_emotions}
\end{table}
The dataset has imbalanced emotion classes, with joy and sadness accounting for 65\% of the samples, anger and fear 26\%, and the remaining 9\% surprise and neutral. 
\revision{To validate the model's performance on our dataset, we randomly sample 250 posts, and two student authors of this paper annotate them by assigning one of six labels. 
We then evaluate the model on 177 posts that both authors agree on as ground-truth-labeled posts and obtain 91.52\% accuracy.}

\descr{Inference.} We use the fine-tuned model to infer the emotions expressed in each post of our dataset, including submissions and comments (note that we combine the text in the title with the body of the submission).
To comply with the training requirements of RoBERTa~\cite{liu2019roberta}, %
we truncate the input text to 512 tokens.
We also remove URLs and convert the text to lowercase. %
Overall, we perform inference on 110M posts, obtaining the following class distributions: Joy (\numprint{7827558}), Sadness (\numprint{2023545}), Anger (\numprint{2939037}), Fear (\numprint{1201658}), Surprise (\numprint{2142924}), Neutral (\numprint{94376186}).

\descr{Results.} Table~\ref{tab:subreddit_emotions} provides an overview of the results on the top 15 subreddits, reporting the number of posts inferred by the emotion detection model and the corresponding average score for each class. 
We observe that /r/NFTsMarketplace has the lowest sadness, anger, and fear scores, while /r/NFT the highest for joy. %
Arguably, the latter is a general subreddit involving users with a generic interest in NFTs, while the former is more centered around buying, selling, and promoting NFTs and, thus, more exposed to gains/losses. mtje

We also find that /r/dogecoin and /r/SHIBArmy have, respectively, the highest and lowest scores for surprise, while /r/Buttcoin has the highest for sadness and anger and the lowest for joy. 
Overall, /r/BitcoinMarkets exhibits the highest level of fear. %
Manual inspection of the subreddit, reveals posts like {\em ``This feels like some serious blood. Who the fuck is buying? I want names. I'm too scared to go all in, but I DID purchase. When we hit 65k in a few weeks I wanna read this thread again.''}

\subsection{Temporal Evolution} %
\label{sec:evolution_emo}

In Figure~\ref{fig:emo_timeline}, we plot the average score of each emotion (anger, fear, joy, sadness, surprise) computed over 1-week timespans to measure how emotions change over time. %
Additionally, we denote specific months with gray regions, aligning them with key events. %

\descr{2011--2015.} With the emergence of cryptocurrencies, communities show mixed emotions. 
In November 2013, as Bitcoin crosses the \$1K mark for the first time, we find a substantial increase in joy and surprise compared to fear, anger, and sadness. 
However, as Bitcoin's price drops below \$180 in January 2015, there is a notable decline in joy and surprise, transitioning from a brief period of joy to one of fear.

\descr{2016--2020.} 
In early 2016, Bitcoin's price starts to increase drastically, reaching a peak of around \$19K by the end of 2017. 
Here we do not find any distinguishable differences in emotions. 
However, in 2018, there is a considerable rise in anger as the cryptocurrency market collapses, an event known as the ``Great Crypto Crash''~\cite{crypto_dotcom_crash} with prices dropping 80\% from January to September 2018.
The trends of joy and sadness remain stable over 2017--2019, despite the cryptocurrency crash.

In March 2020, with widespread lockdowns caused by COVID-19, the cryptocurrency market plunges again, and we observe a surge in emotions of fear, anger, and sadness. 
However, a Bitcoin rally towards the end of 2020, with the price reaching a then-all-time-high of \$42K on January 8, 2021, sees a gradual increase in joy. 
Interestingly, we do not see a decrease in anger but rather a temporary surge in anger. %
Overall, sudden changes in Bitcoin's value seem associated with anger and concern among cryptocurrency communities.

\descr{2021--2022.} %
In early 2021, Bitcoin's value keeps going up, crossing the \$60K mark in March 2021 and reaching an overall all-time high of \$68K in November. 
This coincides with a substantial rise in joy, with a notable decrease in anger and sadness. 
Any subsequent changes in the price of Bitcoin from this point on yield detectable changes in emotions. 
Some other notable events include Coinbase going public on the NASDAQ~\cite{coinbase_public} and other coin prices rising sharply in early 2021, with Dogecoin increasing to \numprint{14000}\% over the first five months of 2021~\cite{doge_joke}.

With the peak in Bitcoin's price in November 2021, we observe a corresponding rise in joy.
However, starting in mid-2022, the price of Bitcoin collapsed, and we see correlated increases in fear, anger, and sadness.

\subsection{Cryptocurrency-related Events}

\begin{figure}[t!]
  \centering
    \includegraphics[width=0.99\columnwidth]{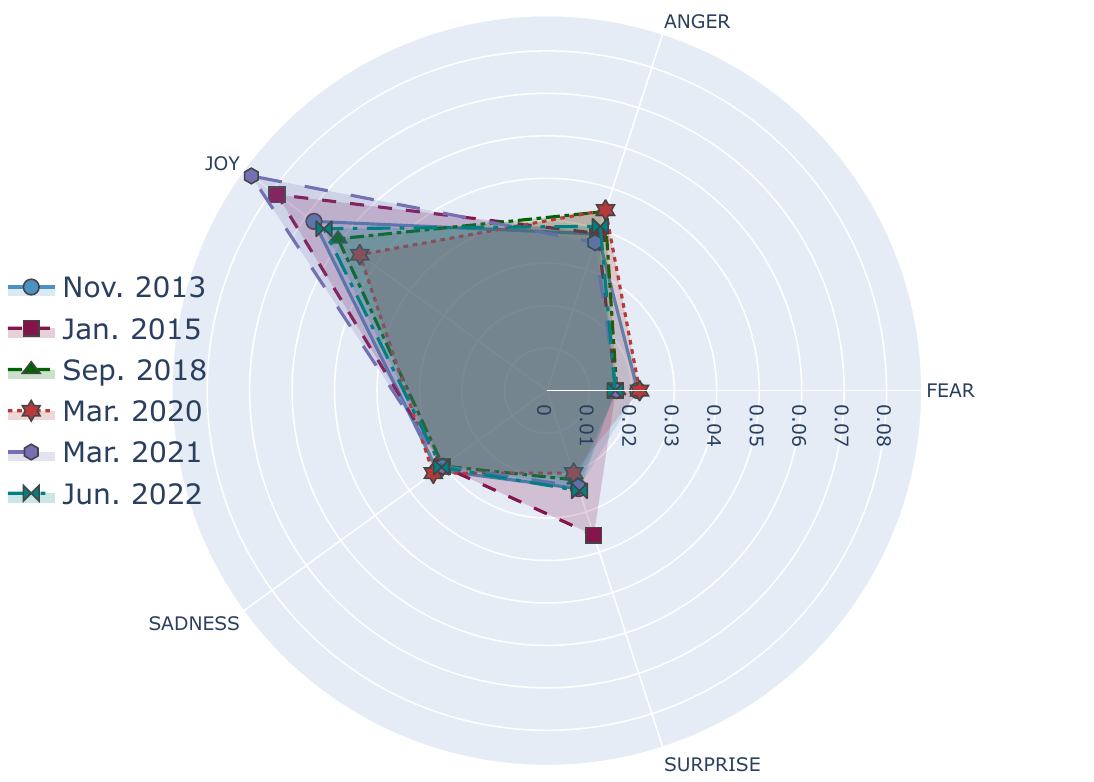}
  \caption{Radar plot with six selected cryptocurrency-related events, with each attribute representing the average emotion scores observed during the respective event month.}
\label{fig:emo_radar}
\end{figure}

\revision{As our emotion detection model captures notable shifts during major cryptocurrency-related events, we now set out to dig deeper into how emotions accompany these events as they unfold. }
\revision{To do so, we select six cryptocurrency-related key events: %
1) Bitcoin surpassing \$1K in November 2013, 2) Bitcoin dropping below \$180 in January 2015, 3) the Great Crypto Crash in September 2018, 4) the COVID-19 pandemic onset in March 2020, 5) Bitcoin crossing \$60K in March 2021, and 6) a 37\% decline in crypto fund assets under management (AUM) in June 2022~\cite{brutal_june22}.}

\revision{We calculate the average score of each emotion over a month when specific events occur and present it as a Radar plot in Figure~\ref{fig:emo_radar}.
The first-ever drastic Bitcoin price decline in January 2015 comes with surprise as opposed to negative emotions like sadness and anger. 
We also observe the highest intensity of anger, fear, and sadness during the onset of the pandemic.}

\descr{Take-aways.} \revision{First, we observe that emotions vary between inter-community and intra-community settings, with distinct differences among communities.
For example, the /r/dogecoin community is the most joyful within its community but experiences the highest intensity of surprise among the top 15 communities (see Table~\ref{tab:subreddit_emotions}).
Second, we reveal that emotions are correlated with cryptocurrency related events.
For example, early in the ecosystem's evolution we see a noticeable shift in joy and surprise.
}

\section{Discussion and Conclusion}
\label{sec:conclusions}
This paper presented a large-scale study of cryptocurrency communities on Reddit.
We compiled a list of 122 cryptocurrency-related subreddits and collected a dataset of 130.8M submissions and comments made on them since their creation.
\revision{(To our knowledge, this is the largest dataset of cryptocurrency communities in terms of both volume and timeline; we will make it available upon final version of the paper.)}

We presented a general characterization of the evolution of the cryptocurrency communities on Reddit, both in terms of user activity and popularity.
Our analysis points to a surge in cryptocurrency-related activity on Reddit within 2021, characterized by heightened excitement and enthusiasm, followed by a sharp decrease in activity at the end of the year, which aligns with the rise and fall of several cryptocurrencies' value. 
We also used temporal analysis and statistical modeling to examine the cross-correlation between cryptocurrency-related online activity and relevant price movements.
Notably, our findings reveal a strong relationship between them, especially for Ethereum, Monero, Cardano, and Polygon, with the changes in online activity mostly preceding price changes.
Finally, we examined the emotional dynamics of the cryptocurrency communities and explored the interplay between emotions and market behavior using an emotion recognition classifier.
\revision{Favorable events within communities, such as upward market performance, are correlated with an increase in joy, while market declines are correlated with increased anger.}
\descr{Broader Perspectives.}  %
\revision{Although we are primarily interested in understanding the cryptocurrency ecosystem with respect to social media activity, the fact that our simple trading strategy resulted in profit indicates a potentially unhealthy relationship between price and social media activity that regulators might want to further explore.}
Possible \emph{negative outcomes} of our work include misrepresentation of the results from users and communities we study~\cite{yudhoatmojo2023understanding}, e.g., with vested interests in the success or failure of cryptocurrencies.
To that end, we paid additional care with language aiming to reduce potential misuse.
Also, recent discussions about the opportunity of using social network data for training AI tools like ChatGPT~\cite{chatgpt} might raise concerns with respect to Reddit's ToS; however, Reddit's changes were only announced on April 18th, 2022, well after our data collection ended (August 2022).
Finally, we note that none of the authors has any {\em competing interests.}

\descr{Limitations \& Future Work.} Our analysis is limited to Reddit; future work should extend to additional social networks and instant-messaging apps that are popular with cryptocurrency enthusiasts, e.g., Twitter, Discord, Telegram, etc.
Moreover, while our analysis reveals strong relationships between cryptocurrency prices and posting activity, it does not imply direct causation.
Market dynamics, news events, and investor behavior also significantly influence both variables. 
Further analysis of these factors will be crucial for a more comprehensive understanding. %

Also, our work is constrained by the fine-tuned RoBERTa model, which requires truncating all posts to 512 tokens; however, this is a general limitation of most of the available transformer-based models. 
\revision{We will improve our emotion detection model by making it a multi-label classifier, which is a known challenge previously discussed in~\citet{nandwani2021review, zad2021emotion}.}

To ease presentation, we omit the full set of results and plots for all 50 cryptocurrencies and the 122 subreddits, as well as additional analyses (e.g., temporal patterns, URL analysis, etc.) -- we plan to make them available with an extended version of this paper.
Finally, we plan to study various important events that affected the cryptocurrency market (i.e., scams), focusing on the toxicity and suicidal ideation levels during and after such important events.

\small
\bibliographystyle{apalike}
%\bibliography{refs}

\end{document}